\documentclass[prc,preprint,showpacs,floatfix,nofootinbib,tightenlines]{revtex4}

\usepackage{amsmath,amssymb,graphicx,bm}

\newcommand{\beqn}{\begin{equation}}
\newcommand{\eeqn}{\end{equation}}
\newcommand{\bea}{\begin{eqnarray}}
\newcommand{\eea}{\end{eqnarray}}

\newcommand{\abinit}{{\it ab initio }}

\newcommand{\fmi}{\, \text{fm}^{-1}}

\newcommand{\la}{\langle}
\newcommand{\ra}{\rangle}

\newcommand{\adag}{a^\dagger}

\newcommand{\nmax}{$N_{\rm max}$}
\newcommand{\hw}{$\hbar\Omega$}
\newcommand{\atw}{$N_{\rm A2max}$}
\newcommand{\ath}{$N_{\rm A3max}$}

\newcommand{\vbt}{\overline{V}^{(2)}_s}
\newcommand{\vbtr}{\overline{V}^{(3)}_s}

\begin{document}


\title{Evolving Nuclear Many-Body Forces with \\ 
		the Similarity Renormalization Group}

\author{E.D.\ Jurgenson}
\email{jurgenson2@llnl.gov}
\affiliation{Lawrence Livermore National Laboratory, P.O. Box
808, L-414, Livermore, CA\ 94551, USA}

\author{P.\ Navr\'atil}
\email{navratil@triumf.ca}
\affiliation{Lawrence Livermore National Laboratory, P.O. Box
808, L-414, Livermore, CA\ 94551, USA}
\affiliation{TRIUMF, 4004 Westbrook Mall, Vancouver, BC, V6T 2A3, Canada}

\author{R.J.\ Furnstahl}
\email{furnstahl.1@osu.edu}
\affiliation{Department of Physics, The Ohio State University, 
Columbus, OH\ 43210, USA}

%
%


\begin{abstract} In recent years, the Similarity Renormalization Group has
provided a powerful and versatile means to soften interactions for \abinit
nuclear calculations. The substantial contribution of both induced and initial
three-body forces to the nuclear interaction has required the consistent
evolution of free-space Hamiltonians in the three-particle space. We present the
most recent progress on this work, extending the calculational capability to the
p-shell nuclei and showing that the hierarchy of induced many-body forces is
consistent with previous estimates. Calculations over a range of the flow
parameter for $^6$Li, including fully evolved NN+3N interactions, show moderate
contributions due to induced four-body forces and display the same improved
convergence properties as in lighter nuclei. A systematic analysis provides
further evidence that the hierarchy of many-body forces is preserved.  
\smallskip

\noindent UCRL--LLNL--462328
\end{abstract}
\smallskip

\maketitle

\section{Introduction}
\label{sec:intro}

A major goal of nuclear structure theory is to make quantitative calculations of
low-energy nuclear observables starting from microscopic inter-nucleon forces. 
Renormalization group (RG) methods can be used to soften the short-range
repulsion and tensor components of available initial interactions so that
convergence of nuclear structure calculations is greatly
accelerated~\cite{Bogner:2006vp,Bogner:2007rx}.  A major complication is that these
transformations change the short-range many-body forces. In fact, any softening
transformation will induce many-body interactions in the course of renormalizing
the matrix elements in a lower sector. To account for these changes, we must
include consistently evolved three-body (and possibly higher) forces in structure
calculations. 

A previous letter~\cite{Jurgenson:2009qs} presented the first such evolution of
three-body forces in free space by using the Similarity Renormalization Group 
(SRG)~\cite{Glazek:1993rc,Wegner:1994,Kehrein:2006,Bogner:2006srg,Bogner:2007srg,Roth:2008km}.
The SRG offers an approach to evolving many-body forces that is technically
simpler than other unitary RG formulations. Irrespective of the chosen initial
Hamiltonian, the evolution produces a variational Hamiltonian and enables smooth
extrapolation of results, in contrast to Lee-Suzuki~\cite{LS1} type transformations which
produce results that are model-space dependent (in both \nmax\ and
$A$)~\cite{NCSMC12}. While
the SRG induces many-body forces as a product of renormalization, these terms
come in a hierarchy of decreasing strength if a hierarchy is
initially present. Particularly useful in an analysis of such a hierarchy are
chiral effective field theories ($\chi$EFTs), which provide a systematic
construction of many-body forces as the initial input to our evolution
calculations.~\cite{Epelbaum:2008ga}. Our results expand on prior evidence that
the SRG explicitly preserves the initial EFT many-body hierarchy as it improves
convergence properties of evolved Hamiltonians.

Section \ref{sec:background} reviews some background material on how the SRG is
applied in these calculations. In Section~\ref{sec:convergence} we explore the
convergence properties of the renormalized Hamiltonians, including new $A$=6
calculations. In Section~\ref{sec:evolution} we present the calculations as a
function of the evolution parameter, and explore the effect of SRG flow on other
initial interactions. Section \ref{sec:hierarchy} dives deeper into the analysis
of how the SRG acts to evolve the input interaction, expanding upon the analysis
done for one-dimensional models~\cite{Jurgenson:2008jp}. We make a brief advertisement of
operator evolution and conclude with comments on the future use of this
approach.

\section{Background}
\label{sec:background}

As implemented in Refs~\cite{Bogner:2006srg,Bogner:2007srg} for nuclear physics,
the SRG is a series of unitary transformations, $U_\lambda$, of the free-space
Hamiltonian, 
\beqn
H_{\lambda} = U_{\lambda} H_{\lambda=\infty}U_{\lambda}^{\dagger} \;,
\label{eq:Hs}
\eeqn
labeled by a momentum parameter $\lambda$ that runs  from $\infty$ toward zero,
which keeps track of the sequence of Hamiltonians ($s = 1/\lambda^4$  is also
used elsewhere~\cite{Bogner:2006srg,Bogner:2007srg}). These transformations are
implemented as a flow equation in $\lambda$ (in units where $\hbar^2 = M = 1$),
\beqn
 \frac{dH_{\lambda}}{d\lambda} = -\frac{4}{\lambda^5} [[T,H_{\lambda}],H_{\lambda}]  \;,
 \label{eq:flow}
\eeqn
whose form guarantees that the $H_\lambda$'s are unitarily 
equivalent~\cite{Kehrein:2006,Bogner:2006srg}. Once the Hamiltonian has been
evolved we also have the option to build the unitary transformation operator
directly as a sum over outer products of the evolved and unevolved
wavefunctions:
\beqn
U_\lambda = \sum_{\alpha} |\psi_\alpha(\lambda)\ra \la \psi_\alpha(0)| \;,
\label{eq:u_lam}
\eeqn
where $\alpha$ is an index over the states in the chosen configuration space.
This feature is useful in applications to external
operators~\cite{Anderson:2010aq}. Note that $U_\lambda$ can also be evolved
directly and the choice of method is open to efficiency and convenience for a
particular use.


The appearance of the nucleon kinetic energy $T$ in Eq.~\eqref{eq:flow} leads to
high- and low-momentum parts of  $H_{\lambda}$ being decoupled, which means
softer and more convergent potentials~\cite{Jurgenson:2007td}. This is evident
in a partial-wave momentum basis, where matrix elements $\la k | H_\lambda | k'
\ra$  connecting states with kinetic energies differing by more than
$\lambda^2$ are suppressed by $e^{-(k^2-k'{}^2)^2/\lambda^4}$ factors and
therefore the states decouple as $\lambda$ decreases. However, decoupling also results
from replacing $T$ in Eq.~\eqref{eq:flow} with other 
operators~\cite{Kehrein:2006,Bogner:2006srg,Glazek:2008pg,Anderson:2008mu}.
The optimal range for $\lambda$ is not yet established and also depends on the
system. Previous experience with SRG and other low-momentum potentials suggested
that running to about $\lambda = 2.0 \fmi$ is a good compromise
between improved convergence from decoupling and the growth of induced many-body
interactions~\cite{Jurgenson:2007td}. Current results show that this limit might
be extended as far as $\lambda = 1.0 \fmi$, at least for lighter nuclei. 


One formal way to see how the two-, three-, and higher-body potentials evolve is
to decompose $H_\lambda$ in second-quantized form~\cite{Kehrein:2006}. We can
write a general $A$-body Hamiltonian as:
\beqn
  H_{\lambda} = \sum_{ij} T_{ij} \adag_i a_j  + 
  \frac{1}{2!^2} \sum_{ijkl} V_{ijkl,\lambda}^{(2)} \adag_i\adag_j a_l a_k
   + \frac{1}{3!^2} \sum_{ijklmn}  V_{ijklmn,\lambda}^{(3)}
   \adag_i\adag_j\adag_k a_n a_m a_l + \cdots 
	\;,
\label{eq:2ndquant}
\eeqn
where $\adag_i$ and $a_i$ are creation and destruction operators with respect to
the vacuum in some single-particle momentum basis. The quantities $T_{ij}$,
$V_{ijkl}^{(2)}$, and $V_{ijklmn}^{(3)}$ denote matrix elements of their
respective operators. Equation~\eqref{eq:2ndquant} \emph{defines} $T_{ij}$,
$V_{ijkl,\lambda}^{(2)}$, $V_{ijklmn,\lambda}^{(3)}$, \ldots as the one-body,
two-body, three-body, \ldots matrix elements at each $\lambda$. By evaluating
the commutators in Eq.~\eqref{eq:flow} using $H_\lambda$ from
Eq.~\eqref{eq:2ndquant}, and normal ordering the resulting terms of
creation/annihilation operators, we find that higher-body potentials are
generated with each step in $\lambda$, even if initially there are only two-body
potentials. We note that in this paper we are not actually evolving in a
single-particle basis as indicated in Eq.~\eqref{eq:2ndquant}, but nothing
\emph{a priori} prevents it as a choice of basis. In particular, the
center-of-mass solutions will factor out in the properly truncated Harmonic
Oscillator single-particle basis. Furthermore the SRG will not mix different
center-of-mass solutions since $T_{CM}$ commutes with the Hamiltonian.


Here we are normal ordering with respect to the vacuum, as opposed to the
in-medium SRG which normal orders with respect to a non-vacuum reference state.
With in-medium normal ordering, SRG evolution generates an $A$-dependent
rearrangement of the higher-body contributions to the evolved interaction; the
density-dependent 0-, 1-, and 2-body normal-ordered interactions are found to
absorb the dominant free-space many-body interactions~\cite{Bogner:2009bt}. For
free-space normal ordering, matrix elements in a given sector are determined
completely by evolution in that sector. In addition, each $A$-body sector
contains as a subset the $(A-1)$-body sector evolutions. Thus, when applied in
an $A$-body subspace, the SRG will ``induce'' $A$-body forces, with $\la T \ra$
fixed, $\la V_\lambda^{(2)} \ra$ determined completely in the $A=2$ subspace
with no dependence on $\la V_\lambda^{(3)} \ra$, $\la V_\lambda^{(3)} \ra$
determined in $A=3$ given $\la V_\lambda^{(2)} \ra$ and $\la V_{\lambda=0}^{(3)}
\ra$, and so on.


Because only the Hamiltonian enters the SRG evolution equations, there are no
difficulties from having to solve T matrices (of the Lippmann-Schwinger
equation) in all channels for different $A$-body systems~\cite{Bogner:2006vp}. However, in a momentum
basis the presence of spectator nucleons requires solving separate equations for
each set of $\la V^{(n)}_\lambda \ra$ matrix elements. In
Ref.~\cite{Bogner:2007qb}, a diagrammatic approach was introduced to handle this
decomposition. But while it is natural to solve Eq.~\eqref{eq:flow} in momentum
representation, it is an operator equation allowing us to use any convenient
basis. Here we evolve in a \emph{discrete} basis of Jacobi-coordinate harmonic oscillator
wave functions, where spectators are handled without a decomposition and induced
many-body interaction matrix elements can be directly identified. Having chosen
a basis, we obtain coupled first-order differential equations for the matrix
elements of the flowing Hamiltonian $H_\lambda$, where the right side of
Eq.~\eqref{eq:flow} is evaluated using simple matrix multiplications.


 The procedures used here build directly on Ref.~\cite{Jurgenson:2008jp}, which
presents a one-dimensional implementation of our approach along with a general
analysis of the evolving many-body hierarchy. We start by evolving $H_\lambda$
in the $A=2$ subsystem, which completely fixes the two-body matrix elements $\la
V_\lambda^{(2)}\ra$. Next, by evolving $H_\lambda$ in the $A=3$ subsystem we
determine the combined two-plus-three-body matrix elements. We can isolate the
three-body matrix elements by subtracting the evolved  $\la V_\lambda^{(2)}\ra$
elements in the $A=3$ basis~\cite{Jurgenson:2008jp}. Having obtained the
separate NN and NNN matrix elements, \emph{we can apply them unchanged to any
nucleus}.  We are also free to include any initial three-nucleon force in the
initial Hamiltonian without changing the procedure. If applied to $A \geq 4$,
four-body (and higher) forces will not be included and so the transformations
will be only approximately unitary. The questions to be addressed are whether
the decreasing hierarchy of many-body forces is maintained and whether the
induced four-body contribution is unnaturally large. We summarize in
Table~\ref{tab:one} the different calculations to be made here for  $^3$H, $^4$He,
and $^6$Li to confront these questions. These calculations will also be made
when other nuclei are considered.

\begin{table*}[tbh-]
\caption{\label{tab:one}Definitions of the various calculations.}
\begin{tabular}{rl}
\hline
\hline
  NN-only | & No initial NNN interaction    \\
   & and do not keep NNN-induced interaction.   \\
  NN + NNN-induced | & No initial NNN interaction   \\
   &    but keep the SRG-induced NNN interaction.   \\
  NN + NNN | & Include an initial NNN interaction    \\
   &  	\emph{and} keep the SRG-induced NNN interaction.   \\
\hline
\hline
\end{tabular}
\end{table*}

%
%

Hamiltonians obtained via free-space SRG evolution are independent of the basis
choice. Up to truncations induced by conversion to a particular basis, a
Hamiltonian evolved to a given $\lambda$ reproduces the results of a Hamiltonian
evolved to the same $\lambda$ in a different basis. Two types of truncations, in
model-space size and $A$, are relevant to controlling the quality and
consistency of SRG evolved interactions. Our calculations are performed in the
Jacobi coordinate harmonic oscillator (HO) basis of the No-Core Shell Model
(NCSM)~\cite{Navratil:2009ut}.  This is a translationally invariant,
anti-symmetric basis for each $A$-body sector, in which a complete set of states
in the model space defines  the maximum excitation of \nmax\hw\ above the
minimum energy configuration, where $\Omega$ is the harmonic oscillator
parameter. Hamiltonians are derived and evolved in this basis and then switched
to a Slater determinant basis. The Jacobi coordinates used to build this basis
have a convenient normalization that treats all $A$-body clusters on an equal
footing. Operators in an $A$-body space, like the $A$-body Hamiltonian, can be
embedded in an $(A+n)$-body space in a straightforward manner. Due to the
antisymmetric nature of the basis, they need only be multiplied by a
combinatoric factor: 
\beqn
{A+n \choose A} = \frac{(A+n)!}{A!(n)!} \;.
\eeqn
For example, a 3-body system has ${3 \choose 2} = 3$ pairs, a 4-body system has
${4 \choose 2} = 6$ pairs and ${4 \choose 3} = 4$ triplets, etc. This embedding
factor was a direct predictor of behavior in one
dimension~\cite{Jurgenson:2008jp}, but in the realistic case, many physical
constraints may complicate the end results including, but not limited to, Pauli
blocking, angular momentum selection rules, and cancellations intrinsic to the
initial Hamiltonian.


 A major drawback of the HO basis is its single intrinsic scale, \hw, which is
problematic for systems with multiple scales. However it is a widely used basis
in part because it facilitates the separation of spurious center-of-mass
solutions and vital to the translationally invariant physics of nuclear
structure calculations. To understand the cutoffs inherent in the finite
oscillator basis we can consider that the local maxima in the harmonic
oscillator function are essentially Gaussians modulated by polynomial terms up
to \nmax. These maxima, in the momentum and coordinate space representations,
will be correlated with the high and low-momentum cutoffs respectively. These
cutoffs have the large \nmax\ behavior:~\cite{thesis}
\beqn
\Lambda_{\rm UV} \sim \sqrt{mN_{\rm max}\hbar\Omega} \quad {\rm and} 
		\quad \Lambda_{\rm IR} \sim \sqrt{\frac{m\hbar\Omega}{N_{\rm max}}}\;.
\label{eq:cutoff}
\eeqn
When \hw\ grows large, individual oscillations are large and lose resolution on
the small details in the momentum basis potential that correspond to large $r$
structures. However, high \nmax\ polynomials have many small oscillations at low
momenta compensating for the large \hw\ value. Thus, $\Lambda_{\rm IR}$ is
lowered and $\Lambda_{\rm UV}$ is raised by increasing \nmax\ as expected when
the basis is extended towards completeness. Note that only the value of \nmax,
not \hw, affects the number of matrix elements in the basis, so the
computational cost is the same for each \hw. Changing \hw\ effects the balance
between $\Lambda_{\rm IR}$ and $\Lambda_{\rm UV}$ completeness.

Note that the behavior attributed to $\Lambda_{\rm IR}$ does not manifest as an
explicit cutoff in the momentum representation, but rather a distortion of
matrix elements at low momentum. Specifically, the effective cutoff operator in
momentum representation  displays bands of ringing artifacts along the
off-diagonal direction that ultimately behave as a cutoff; both smaller \hw\ and
larger \nmax\ bases alleviate this effect, as is apparent from
Eq.~\eqref{eq:cutoff}.


Because of computational constraints we were forced to apply separate
truncations, \atw\  and \ath, to the $A = 2$ and $A = 3$ sectors of the initial
Hamiltonian (see Table~\ref{tab:truncs}). In previous
work~\cite{Jurgenson:2009qs} with $^3$H and $^4$He
and an initial $\chi$EFT interaction, we found that \atw\ = \ath\ = 32 was
sufficient because these nuclei are less sensitive to the asymptotic behavior of
the oscillator wavefunctions. However, for $^6$Li or $^4$He using a harder
potential (such as Argonne $V_{18}$),  larger space  was required for the
initial NN Hamiltonian. When needed we used \atw\ = 300, which is more than
enough to accommodate any potential at any relevant \hw. In some calculations we
were restricted to \atw\ = 196, but these cases are also converged to the keV
level. The $A = 3$ basis size grows much more quickly so that the evolution of
Hamiltonians above \ath\ = 40 are very intense computations. In our final
results there is a slight effective truncation of the induced three-body forces,
but this is only a truncation of the initial interaction that is then evolved.
Additionally, the NNN interaction is a perturbative correction to the NN so this
truncation has a small impact on our results. These truncation issues are
addressed throughout the paper when discussing the convergence properties of our
results.

\begin{table*}[tbh-]
\caption{\label{tab:truncs}Definitions of the truncations used on initial
Hamiltonians.}
\begin{tabular}{rl}
\hline
\hline
 \nmax : &  The size of the final A-body space   \\
  \ath : &  The size of the basis for initial three-body matrix elements  \\
  \atw : &  The size of the basis for initial two-body matrix elements  \\
\hline
\hline
\end{tabular}
\end{table*}


The present calculations make use of both the Jacobi oscillator basis described
above and a Slater determinant oscillator basis often referred to as the
$m$-scheme. The size of the Jacobi basis scales well with \nmax\ but poorly with
$A$ due to the effort involved in antisymmetrization. A Slater determinant basis
trades ease of antisymmetrization for very large (dimensions into the billions)
but sparse matrices, solvable with the Lanczos algorithm. Given the convergence
advantages provided by the SRG, we obtain the initial Hamiltonians in the Jacobi
basis, evolve them in $A$=2 and 3, and  transform them into a Slater determinant
basis for use in existing configuration interaction (CI) codes. 

Our calculations are limited by the size of the input three-body interaction
file: the present code is not able to distribute the matrix elements among nodes
and therefore must hold the entire $A$-body Hamiltonian on each node. For our
calculations of $^6$Li at \nmax\ = 8, the 3-body file is 13Gb of matrix elements
in addition to 2-body matrix elements already stored. This is the largest
calculation possible on most available nodes with 16 or 32 Gb of memory; at
\nmax\ = 10 the 3-body file is 33Gb. However, the $m$-scheme code MFDn of Vary
et al.~\cite{mfd_code} is capable of distributing the input matrix elements
among several nodes and efforts are underway to perform these calculations in
larger spaces. Furthermore, the size of these input files can be dramatically
reduced in the future by implementing a compression scheme based on angular
momentum couplings and calculations could be extended with the importance
truncation method~\cite{IT_NCSM,ROTH}. Future calculations of $^8$Be, $^{10}$B,
and $^{12}$C are planned. Recent results of $^{12}$C and $^{16}$O using
importance truncation provide a benchmark for future efforts~\cite{Roth_pc}. 


For the lower $\lambda$'s in the lighter nuclei, our predictions for
ground-state energies are fully converged.  However, in other cases we need to
extrapolate the energies to \nmax$ = \infty$. Here we use the same extrapolation
procedure applied in ~\cite{Bogner:2007rx}. The model used for ground-state energies
is
\beqn
  E_{\alpha i} = E_\infty + A_\alpha\, e^{-b_\alpha N_i}
  \;,
  \label{eq:Ealphai}
\eeqn
where $\alpha$ labels the \hw\  values, $i$ the \nmax\ values for each $\alpha$,
and $A_\alpha$ and $b_\alpha$ are constants. The goal of a fit to the following
calculations is to determine the common parameter $E_\infty$, which is the
estimate for the ground-state energy extrapolated to \nmax$ = \infty$. 

This can be cast as a
one-dimensional constrained minimization problem with the function
\beqn
  g(E_\infty) = \sum_{\alpha,i} 
    ( \log(E_{\alpha i} - E_\infty) - a_\alpha - b_\alpha N_i )^2
    /\sigma_{\alpha i}^2 
    \;,
    \label{eq:residual}
\eeqn
where the $\{a_\alpha\}$ and $\{b_\alpha\}$ are determined directly within the
function $g$ by invoking a constrained linear least-squares minimization
routine. The constraint is the bound  $E_\infty \leqslant \min(\{E_{\alpha
i}\})$, where $E_\infty < 0$ and ``min'' means ``most negative''. We can also
allow for weights depending on \nmax\ and/or \hw.
In the present investigation, we apply Eq.~(\ref{eq:residual}) with only the
\hw\ value that yields the lowest energy in the largest space, weighting
different \nmax\ by the slope of the energy vs.\ \nmax, and using the spread of
results from neighboring \hw\ values to determine a conservative confidence
interval for the extrapolation.  Alternative approaches to extrapolation in the
NCSM are described in Ref.~\cite{Maris:2008ax}.


We have considered a variety of interactions as initial inputs to the SRG
evolution, including chiral EFT, Argonne $V_{18}$~\cite{pot_av18}, and
CD-Bonn~\cite{pot_cdbonn}. The initial ($\lambda = \infty$) chiral NN potential
is the 500\,MeV N$^3$LO interaction from Ref.~\cite{N3LO}. With the chiral
potential we also have available an initial NNN potential at N$^2$LO
~\cite{Epelbaum:2002vt} in the local form of Ref.~\cite{navratil_3N}, with
constants $c_D = -0.2$ and $c_E = -0.205$ fit to the average of triton and
$^3$He binding energies and to triton beta decay as described in
Ref.~\cite{Gazit:2008ma}. NCSM calculations with these initial chiral
interactions and the parameter set in Table~I of Ref.~\cite{Gazit:2008ma} yield
energies of $-8.473(4)\,$MeV for $^3$H and $-28.50(2)\,$MeV for $^4$He compared 
with $-8.482\,$MeV and $-28.296\,$MeV from experiment, respectively. There is a
20\,keV uncertainty in the calculation of $^4$He from incomplete convergence but
a 200\,keV discrepancy with experiment. The latter is consistent with the
omission of three- and four-body chiral interactions at
N$^3$LO~\cite{Rozpedzik:2006yi}. These provide the scale for assessing whether
induced four-body contributions are important compared to other uncertainties.
The best result for $^4$He using the AV18 potential is
$-24.23(1)\,$MeV~\cite{pisa_group} and using CD-Bonn we compare to
$-26.1(1)\,$MeV.  Here there are larger discrepancies with the experimental
values due to the omission of consistent initial three-body interactions, but
these calculations are still useful to assess the effects of induced NNN.

For $^6$Li calculations we use only the chiral interactions at
N$^3$LO~\cite{N3LO} for NN
and at N$^2$LO for NNN in the form described above. The best existing binding
energy with the N$^3$LO interaction, using a Lee-Suzuki based renormalization up
to \nmax\ = 14, is $28.5 \pm 0.5$ MeV. With NNN included the converged value is
$32.5 \pm 0.5$ MeV~\cite{Navratil:2003ib}. The truncations analogous to \atw\
and \ath\ for these calculations were \atw\ = 400 and \ath\ = 40, equivalent to
the initial Hamiltonian inputs for the present work. Results are generally not
dependent on the particular values of the LECs in a range of $c_D \sim -2 \mbox{
to } +2$ (with $c_D$ and $c_E$ constrained by the fit to
$^3$H~\cite{Gazit:2008ma}) but some observables may be particularly sensitive as
discussed in Ref.~\cite{Navratil:2007we}.


\section{Convergence}
\label{sec:convergence}

\begin{figure}[htb]
\includegraphics*[width=8cm]{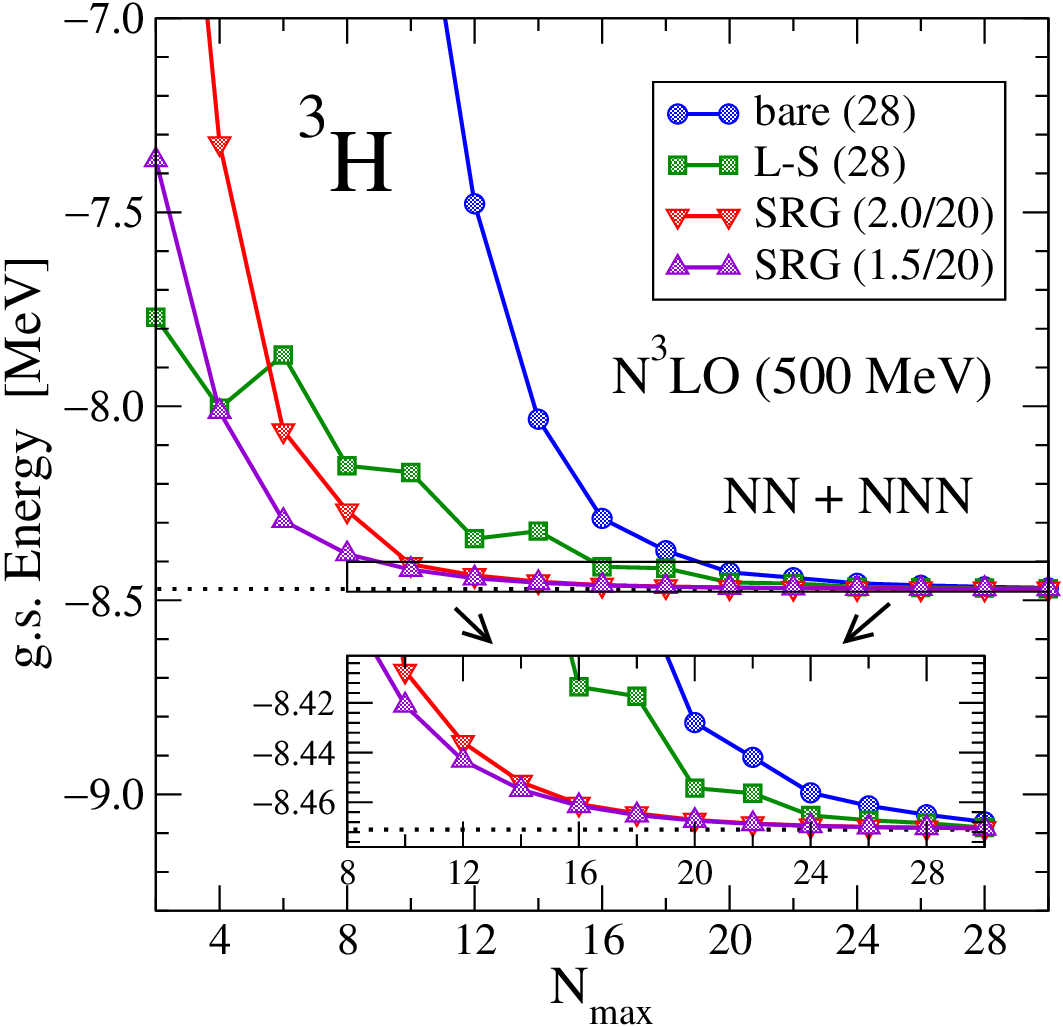}
\caption{(Color online) Ground-state energy of $^3$H as a function of the basis
size \nmax\ for an N$^3$LO NN  interaction~\cite{N3LO} with an
initial NNN interaction~\cite{Epelbaum:2008ga,Gazit:2008ma}.  Unevolved
(``bare'') and Lee-Suzuki (L-S) results with \hw\ = 28 MeV are
compared with SRG evolved to $\lambda =
2.0\fmi$ and $1.5 \fmi$ with \hw\ = 20 MeV.}
\label{fig:h3_convergence}
\end{figure}


In Fig.~\ref{fig:h3_convergence}, we show the triton ground-state energy as a
function of the oscillator basis size, \nmax. The convergence of the bare
interaction is compared with the SRG evolved to $\lambda$ = 2.0 and 1.5 $\fmi$.
The oscillator parameter \hw\ in each case was chosen to optimize the
convergence of each Hamiltonian. We also compare to a Lee-Suzuki (LS)
calculation (green squares), which has been used in the NCSM to greatly improve
convergence~\cite{Nogga:2005hp,Navratil:2007we}. All of these effective
interactions result from unitary transformations. The LS is done within the
model space of a target nucleus, in contrast to the free-space transformation of
the SRG, which yields nucleus-independent matrix elements. Consequently, the LS
results are non-variational independent calculations at each \nmax\ while the
SRG-evolved Hamiltonians can be simply truncated to produce the curves shown. A
dramatic improvement in convergence rate compared to the initial interaction is
seen even though the $\chi$EFT initial interaction is relatively soft. The SRG
acts to decouple high-momentum degrees of freedom so the UV part converges
faster with respect to \nmax. Thus, once
evolved, a much smaller \nmax\ basis is adequate for a particular accuracy.

\begin{figure}[thb]
\includegraphics*[width=8cm]{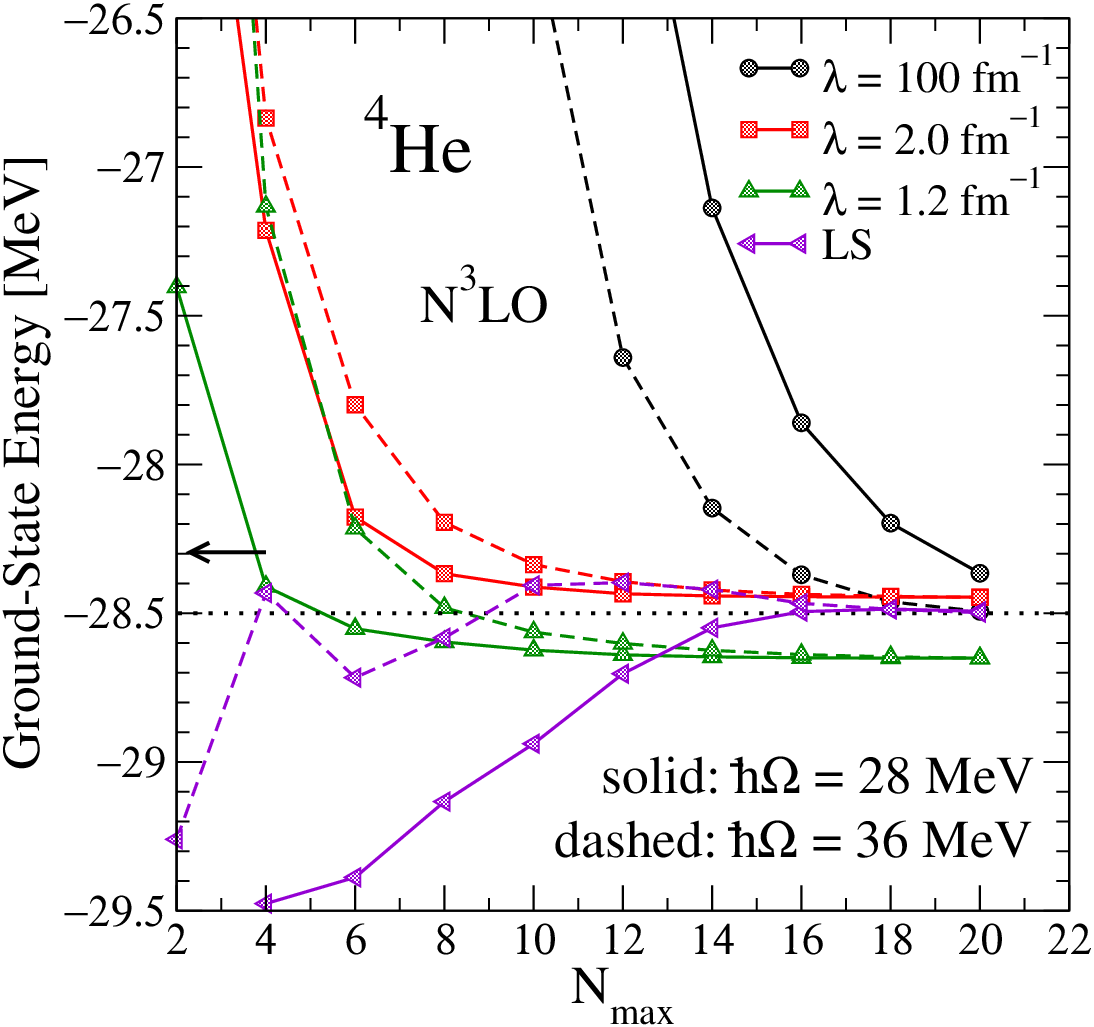}
\caption{(Color online) Ground-state energy of $^4$He as a function of the basis
  size \nmax\ for an N$^3$LO NN  interaction~\cite{N3LO} with an initial NNN
  interaction~\cite{Epelbaum:2008ga,Gazit:2008ma}. Unevolved ($\lambda$) results are
  compared with Lee-Suzuki (L-S) and SRG evolved to $\lambda =
  2.0\,\mbox{and}\,1.2\fmi$  at \hw\ = 28 and 36 MeV. The black arrow indicates
  the experimental value.}
\label{fig:he4_convergence}
\end{figure}


Figure~\ref{fig:he4_convergence} illustrates for $^4$He the same rapid
convergence with \nmax\ of an SRG-evolved interaction.  However, in this case
the asymptotic value of the energy differs slightly because of the omitted
induced four-body contribution. The difference can be as large as 100 keV for
$\lambda = 1.0$ but no larger than 50 keV for the substantially evolved $\lambda
= 2.0 \fmi$. The SRG-evolved asymptotic values for different $\hbar\Omega$
(solid vs. dotted curves) differ by only 10\,keV, so the gap between the
converged bare/L-S result and the SRG result is dominated by the induced NNNN
rather than incomplete convergence. Convergence is even faster for lower
$\lambda$ values, ensuring a useful range for the analysis of few-body systems.
However, because of the strong density dependence of four-nucleon forces, it
will be important to monitor the size of the induced four-body contributions for
heavier nuclei and nuclear matter. In Section~\ref{sec:hierarchy} we present a tool
for analyzing the growth of induced many-body forces.

Also evident in Fig.~\ref{fig:he4_convergence} is the evolving dependence on
\hw. Calculations are variational in \hw\ with the optimal value indicating
balance between $\Lambda_{UV}$ and $\Lambda_{IR}$.  For the initial Hamiltonian,
the limit of \nmax=20 is small and the larger \hw\  is necessary to provide a
sufficient $\Lambda_{UV}$. The larger IR cutoff due to higher \hw\ is less of a
problem than the smaller UV cutoff due to the low \nmax. However, if the initial
Hamiltonian is evolved in the \nmax=32 space, then more UV information is
shifted down into the \nmax=20 space. Now the high IR cutoff is more significant
and a lower \hw\ is more optimal. In the figure, one can see that the \hw\ = 28
calculation (solid curves) has significantly better convergence properties for
lower $\lambda$, especially at the small \nmax\ ($\leq 8$) that is crucial for
larger $A$.

\begin{figure}[thb]
\includegraphics*[width=7.7cm]{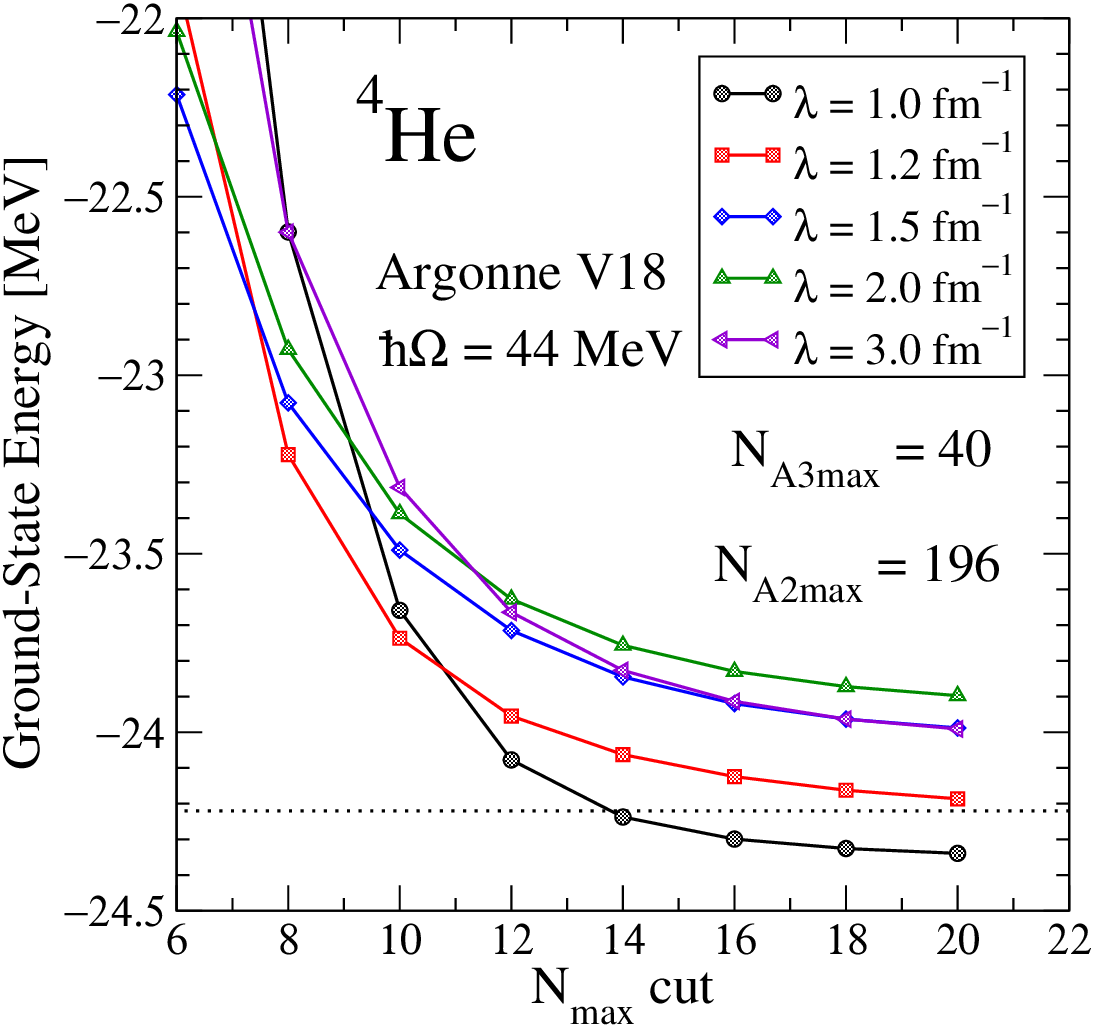}
\includegraphics*[width=7.7cm]{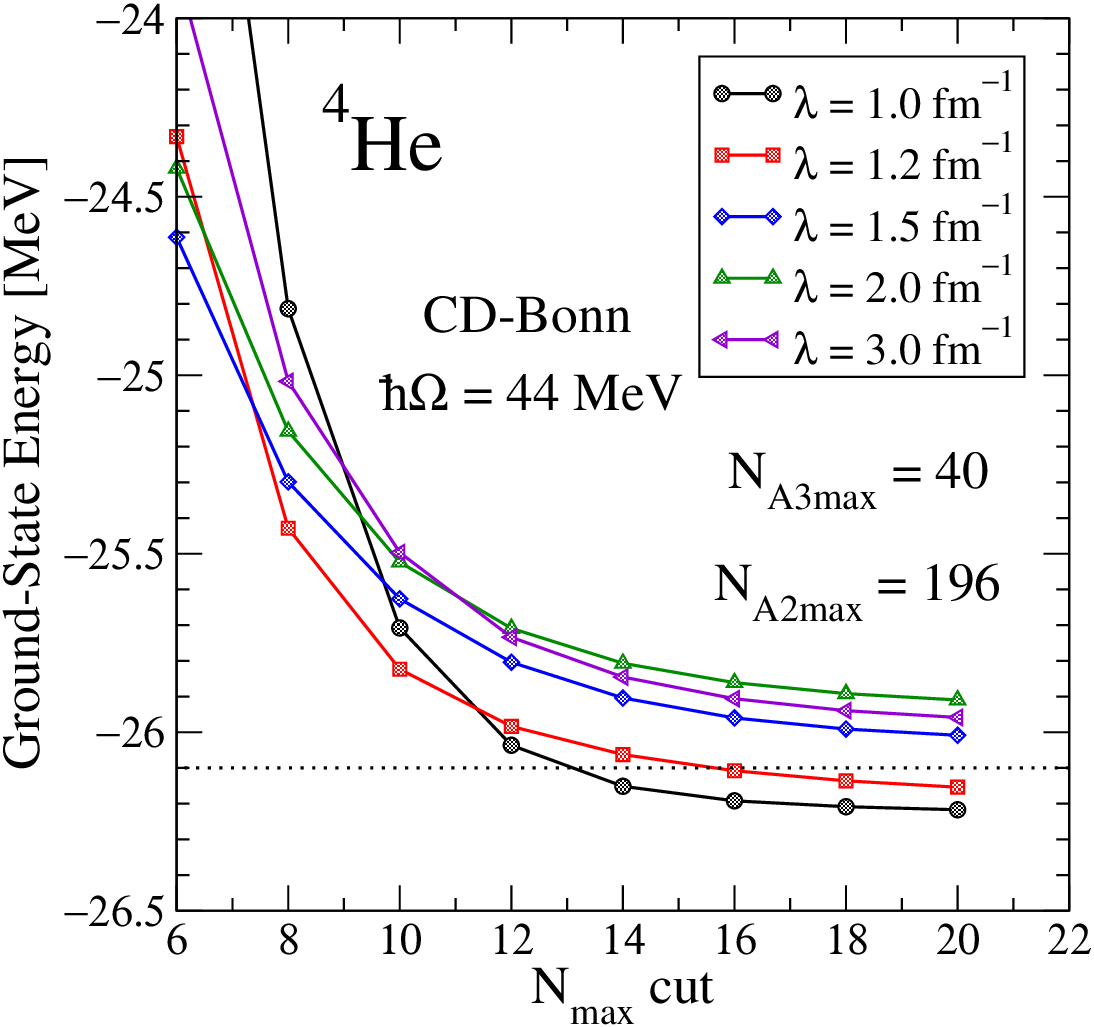}
\caption{(Color online) Ground-state energy of $^4$He for select $\lambda$ as a
function of the basis size \nmax\ for the AV18~\cite{pot_av18} and
CD-Bonn~\cite{pot_cdbonn} interactions. Results
are shown for \hw\ = 44 MeV with \ath\ = 40. Dotted lines indicate current best
results for these potentials~\cite{pisa_group}.}
\label{fig:he4_conv_av18}
\end{figure}


Evolving Hamiltonians such as CD-Bonn and Argonne $V_{18}$ also results in much
improved convergence properties, as seen in Fig.~\ref{fig:he4_conv_av18}. Here
a large initial $A$=2 cutoff, \atw, is crucial due to the strong high-momentum
components in the AV18 potential. However, at \atw=196 the NN-only results are
converged with respect to variation in \atw\ to within 1keV. For CD-Bonn we found
that \ath=40 and \nmax=20 was sufficient to converge results to within 30 keV.
Calculations for the Argonne potential require a bit more effort, obtaining
convergence with respect to \ath\ to within 130 keV. The optimal frequency for
both of these interactions evolved to $\lambda = 2.0$ was found to be \hw\ = 44
MeV. Larger values for \ath\ are possible for these potentials, but the current
level of convergence is sufficient to observe the qualitative behavior of the
SRG in Section~\ref{sec:evolution}.


A multitude of chiral EFT interactions are available for use in initial
Hamiltonians. We have used here the one version (500 MeV N$^3$LO from
Ref.~\cite{N3LO}) for which the accompanying three-nucleon terms have been
rigorously fit to data. An important task in the future will be to apply SRG
techniques to many more available interactions to compare and contrast them.
Some of these can be significantly harder than that chosen here, so running to
low $\lambda$ would be especially important though computationally expensive.

\begin{figure}[thb]
\includegraphics*[width=7.7cm]{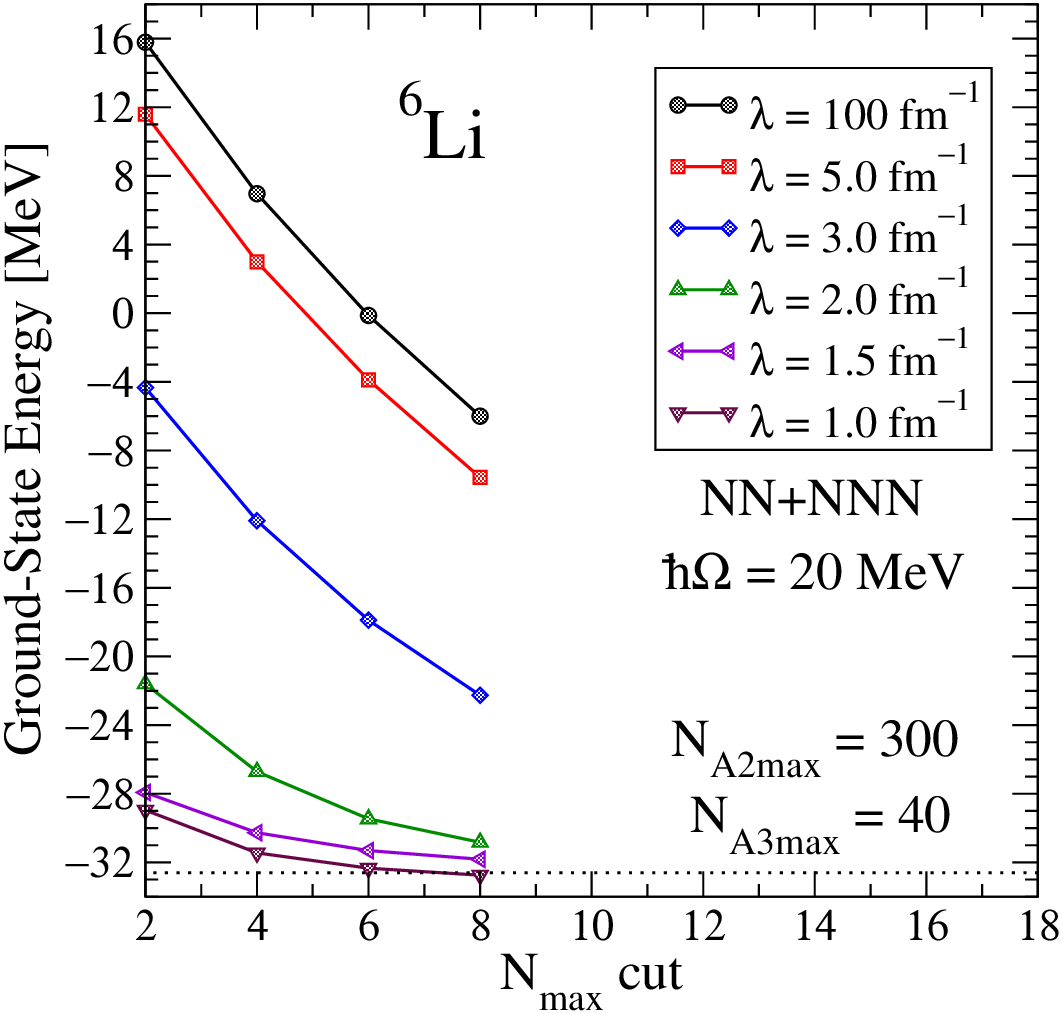}
\caption{(Color online) Ground-state energy of $^6$Li as a function of the basis
 size \nmax\ for an initial N$^3$LO NN interaction~\cite{N3LO} and an initial
 N$^2$LO NNN interaction~\cite{Epelbaum:2008ga,Gazit:2008ma} evolved to various
 $\lambda$. Note the large scale and complete lack of convergence for unsoftened
 Hamiltonians ($\lambda = 100 \fmi$). The dotted line is the best Lee-Suzuki
 result. }
\label{fig:li6_conv_bare}
\end{figure}


Figure \ref{fig:li6_conv_bare} shows just how important a softening
transformation is to achieve convergence in light nuclei. Here we plot $^6$Li
binding energies up to \nmax\ = 8 for several $\lambda$'s from 100 to 1.0
$\fmi$. 
A meaningful extrapolation is simply
not possible, even with the relatively soft chiral potential, without some form
of softening renormalization like the SRG or Lee-Suzuki type transformations. A
key advantage of the SRG program is the ability to perform systematic
extrapolations to spaces that are computationally inaccessible. 

Compare this to the case of $^4$He where the initial chiral EFT Hamiltonian is
sufficiently soft to produce a nearly converged result at \nmax\ = 20. For $A =
6$ we are restricted to smaller \nmax\ because the basis scales with \nmax\ much
faster than at $A = 4$. In addition the radius of $^6$Li is larger and requires
a lower IR cutoff. Thus more oscillator basis states of the initial interaction
are required to accurately describe this nucleus. In other words, even if we
could perform the \nmax\ = 20 calculation of $^6$Li, it would still not converge
as well as $^4$He does at that level. With the SRG the information of the larger
basis can be moved into a smaller space in a smooth controlled way.

\begin{figure}[thb]
\includegraphics*[width=16cm]{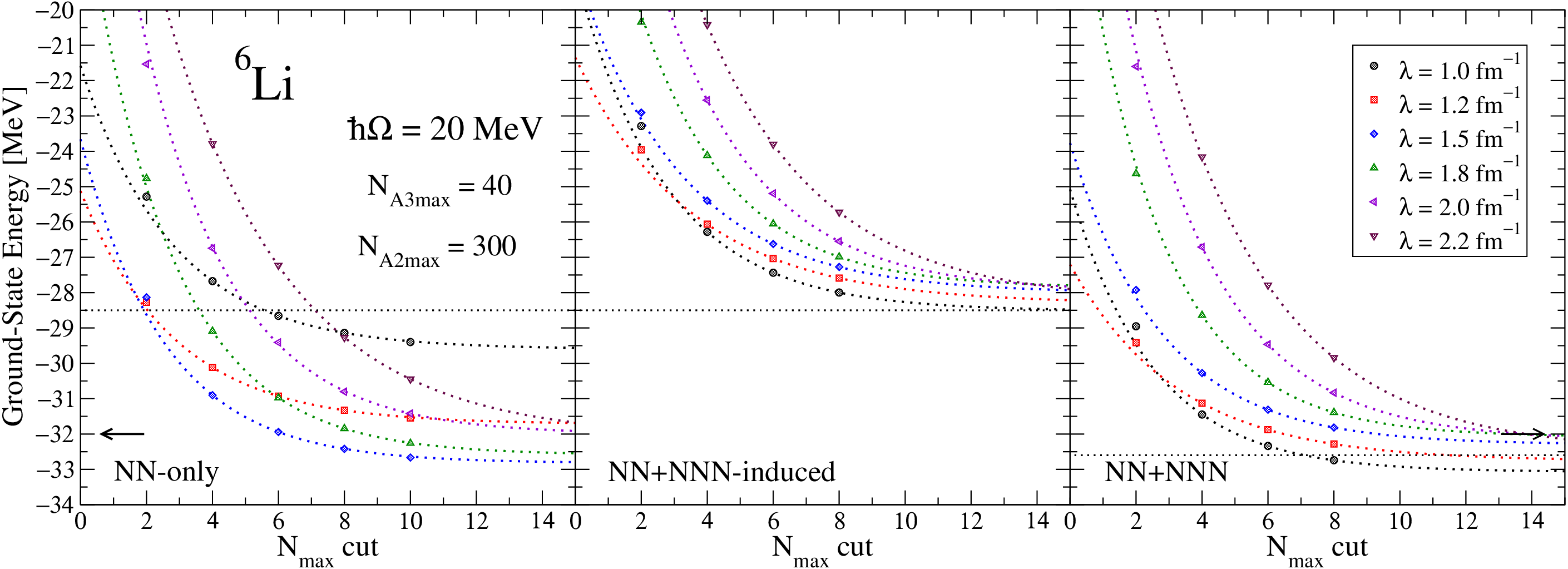}
\caption{(Color online) Ground-state energy of $^6$Li as a function of the basis
 size \nmax\ for an initial N$^3$LO NN interaction~\cite{N3LO} and an initial
 N$^2$LO NNN interaction~\cite{Epelbaum:2008ga,Gazit:2008ma} evolved to various
 $\lambda$. Here the initial NN potential was included up to \atw\ = 300 and the
 NNN up to \ath\ = 40. Results are compared with the best Lee-Suzuki shown
 by the thin black dotted line (see text). The dotted lines extending from the curves are
 examples of the extrapolations made throughout.}
\label{fig:li6_convergence}
\end{figure}


In Fig.~\ref{fig:li6_convergence} we show the convergence patterns of the ground
state of $^6$Li on a more detailed scale. Here the three different calculations
of Table \ref{tab:one} are shown side-by-side for clarity, with several
$\lambda$ values. The initial Hamiltonian was defined by the truncations \atw\ =
300 and \ath\ = 40 and the truncation errors from (or convergence with respect
to) these limits are 1 keV and 80 keV respectively. The $^6$Li calculation was
performed up to \nmax=8 for the three-body matrix element versions, and \nmax=10
for  NN-only (possible with only two-body matrix elements). Further calculations
with three-body matrix elements will require a distributed memory approach like
MFDn of Vary et al~\cite{mfd_code} and the other techniques mentioned above; such
codes will be used for future calculations of larger p-shell nuclei. The
straight dotted line shows the converged Lee-Suzuki results for NN-only and
NN+NNN calculations.

While we can see the improvement in convergence by the data points alone, we can
only measure the effect of induced many-body forces by considering the
extrapolated converged value for each $\lambda$. We provide a sample of the
extrapolations we will use to assess the converged values for $^6$Li, in the
form of the dotted lines extending from each curve. The spread of these lines at
large \nmax\ is the chief indicator of remaining scale dependence in the
results. In the NN-only plot on the left, one can clearly see the large spread
due to omitted induced three-body matrix elements. This spread is
decreased in the other plots by including 3-body matrix elements, first in
the center by including those induced by the renormalization, and on the
right by also including the initial NNN strength. In both NNN-inclusive plots
the curves have a smooth qualitative progression from higher to lower $\lambda$,
indicating less interplay between attractive and repulsive components of the
interaction, as discussed in Section~\ref{sec:hierarchy}.

The size of induced 4- to 6-body forces in this calculation is estimated by
measuring the spread of the lines in the center and right plots, or
alternatively by considering the slope of the binding energy as a function of
lambda. However, the spread is actually smaller than it appears here because
only the $\lambda$=1.5 and 1.8 curves are satisfactorily converged at this
\hw. The curves for other $\lambda$'s are optimal at different values of \hw\ and
their converged values are not accurately represented in this simple example.
However, the full extrapolation procedure does indeed take this \hw\ dependence into
account and this figure serves as a visual reminder of the process.

Finally, we mention the sensitivity of the extrapolations to the range in
\nmax\ used to fit the exponential function. Results at \nmax\ = 10 with NN-only
allow an assessment of the extrapolation procedure. We find that including the
\nmax\ = 2 points bias the extrapolation high, so the best estimates use \nmax\
= 4-8 when NNN matrix elements are included.

\begin{figure}[thb]
\includegraphics*[width=12cm]{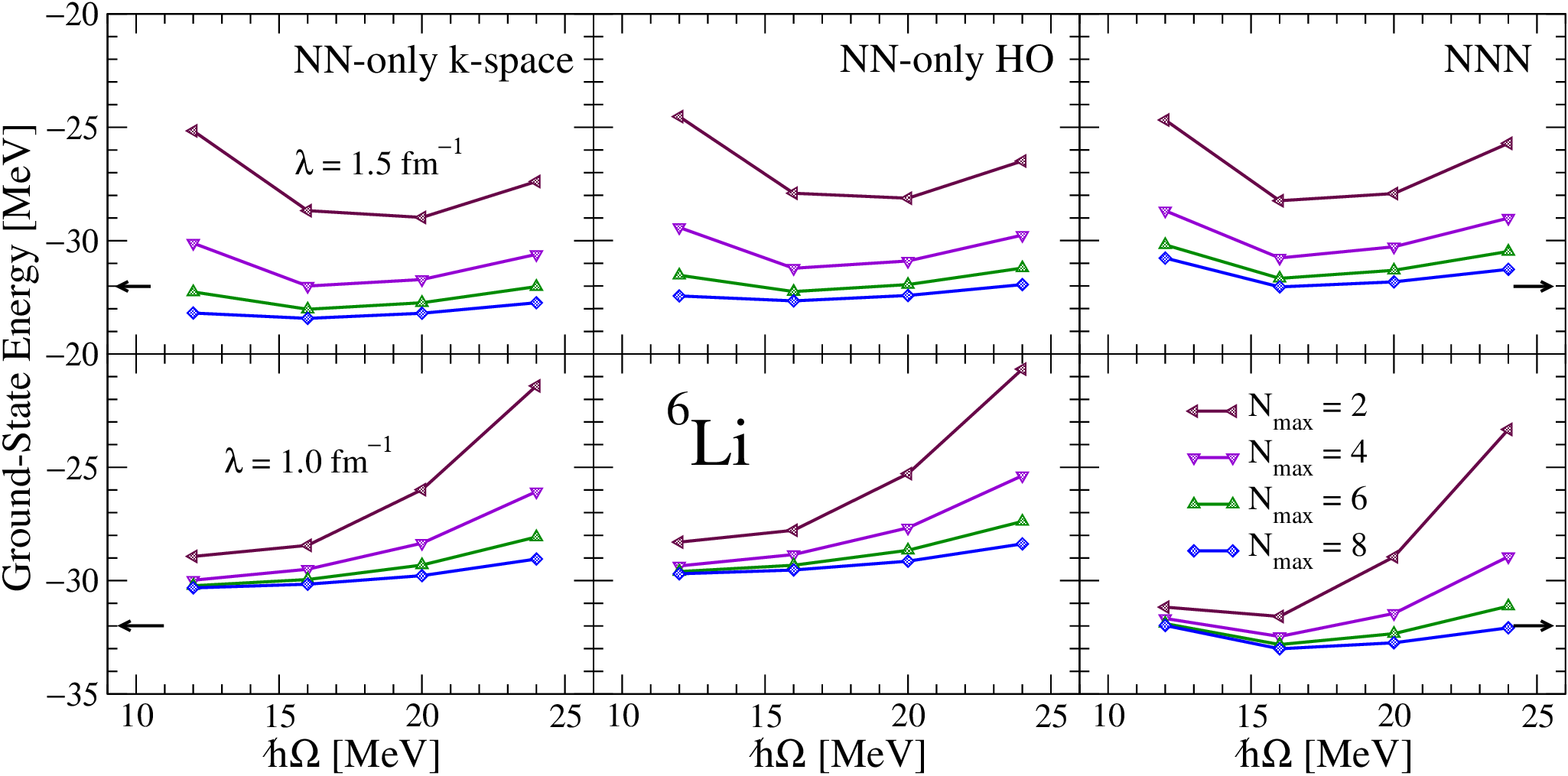}
\caption{(Color online) Ground-state energy of $^6$Li as a function of the
oscillator frequency, \hw. Again we use a N$^3$LO NN~\cite{N3LO} up to \atw\ =
300 and N$^2$LO NNN~\cite{Epelbaum:2008ga,Gazit:2008ma} up to \ath\ = 40. The
left and center panels are NN-only while the right panels include both induced
and initial NNN. The black arrows indicate the experimental value. Note for
comparison, the left  is shifted relative to the center due to the difference in
initial interactions.}
\label{fig:li6_hw}
\end{figure}


Figure \ref{fig:li6_hw} shows the convergence of the $^6$Li ground-state as a
function of \hw\ for selected $\lambda$'s. The separate panels compare, for two
different $\lambda$'s (top and bottom), the current results with a previous
study \cite{Bogner:2007rx} of NN-only calculations where the evolutions were
performed in momentum space. The momentum-space evolutions used only the
neutron-proton part of the interaction, $V_{np}$, as an average for the complete
NN interaction and added the Coulomb contribution separately afterward, causing
a systematic overbinding of about 1 MeV. We have used the whole
N$^3$LO~\cite{N3LO} interaction and included the Coulomb in our evolutions, but
have checked that we recover the previous NN-only results with $V_{np}$. The
left panel shows the momentum-space evolved calculations.  The center panel
shows a reproduction of those results with a systematic shift due to the revised
handling of the initial interaction. The right panel shows the full NN+NNN
calculation. All three panels show good correspondence in \hw\ dependence
between the previous and current calculations, indicating that the NN-only
calculations are good predictors of the minima for the larger 3N-inclusive
version.

As discussed above, a lower \hw\ results in a lower cutoff, $\Lambda_{UV}$, (see
Eq.~\eqref{eq:cutoff}) and requires a larger basis (larger \nmax) to achieve the
same convergence. This can be seen in the NN+NNN panel by observing the trend in
\nmax\ for each \hw. The values that each of the \hw\ spaces are converging
towards are different, indicating that the initial Hamiltonian has been
truncated by different incomplete bases and will not obtain the same result; the
lowest \nmax\ curve is not yet flat. This is especially evident at the smallest
\hw\ which has the worst truncation. Improving this convergence requires
increasing the three-body basis size, \ath. Here, at \ath\ = 40, the dimension
of a single $A=3$ channel (with quantum numbers $J^{\pi}T$) can be 7--8000
states, requiring 60--70 nodes for 12 hours with a hybridized MPI-OpenMP
evolution code. This is currently the most significant computational bottleneck
in our SRG program.

\begin{figure}[htb]
\includegraphics*[height=6.2cm]{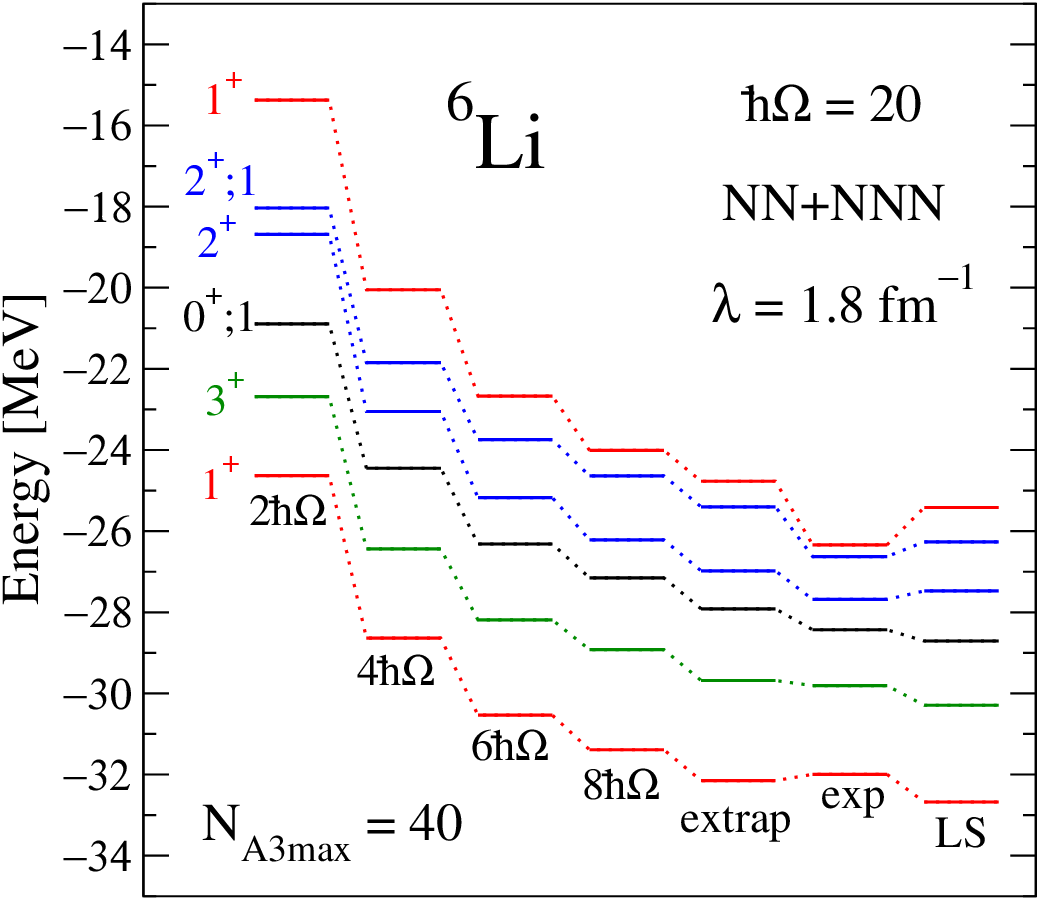}
~\hspace{.2cm}
~\includegraphics*[height=6.2cm]{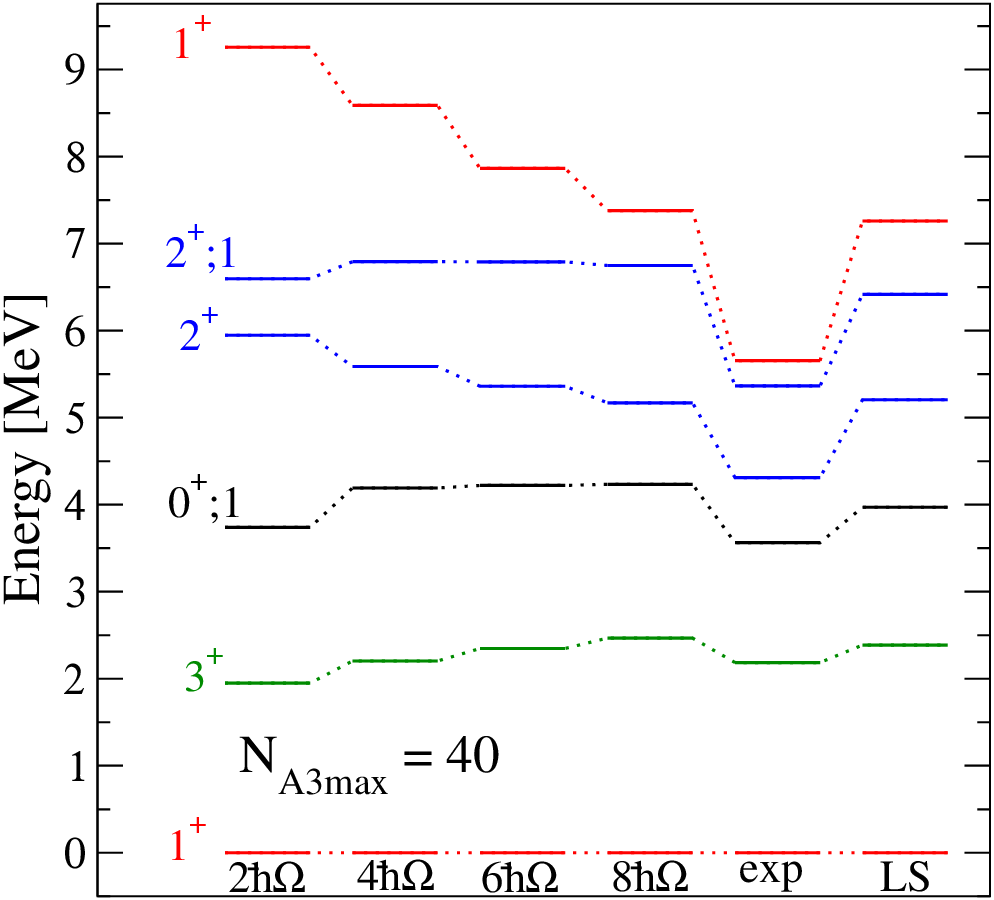}

\vspace{.5cm}
\includegraphics*[height=6.2cm]{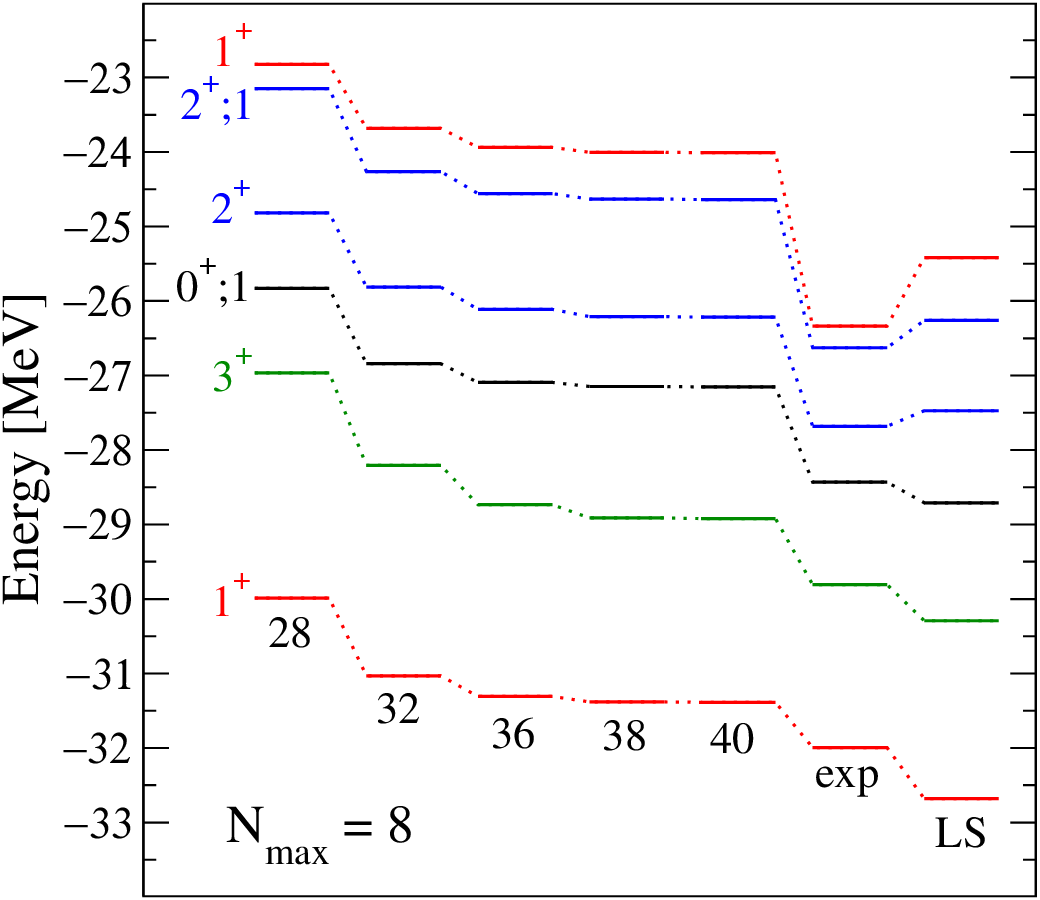}
~\hspace{.2cm}
~\includegraphics*[height=6.2cm]{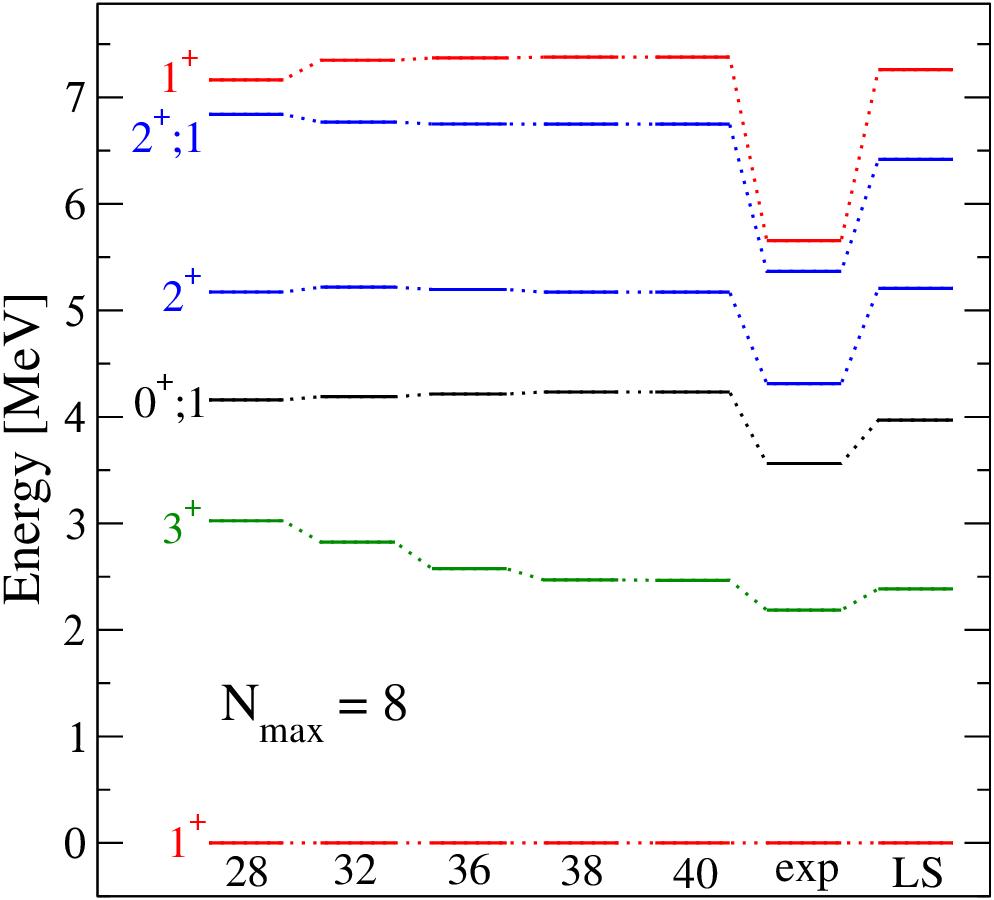}
\caption{(Color online) Spectrum of $^6$Li as a function of \ath. On top
(bottom) shows convergence in \nmax\  (\ath). Left (Right) show absolute
(excitation) energies. All four panels are for $\lambda = 1.8 \fmi$ with NN+NNN at
\hw\ = 20 MeV and \atw\ = 300.  }
\label{fig:Li6_spectra}
\end{figure}


In Fig.~\ref{fig:Li6_spectra}, we show the spectrum for $^6$Li in absolute level
energies on the left and excited state spacings on the right. We chose
$\lambda=1.8 \fmi$ due to sufficient convergence as indicated by our
extrapolation procedure. This example happens to closely match the excitation
spectrum of available LS based results, though the variation in $\lambda$ is not
large. We also include a spectrum with the excitation energies shifted to the
converged ground state energy.

The upper pair of plots shows the convergence with respect to the final \nmax\ 
of the $^6$Li calculation. In the upper left pane we can see a consistent
convergence pattern from 2\hw\ down to 8\hw, but the results are clearly not
converged at 8\hw.  On the right, the excitation energies indicate that the
higher $J$ states are converging more slowly due to stronger dependence on the
higher $A$=3 partial-waves. This was tested by using various levels of
truncation in the initial Hamiltonian for channels with higher values of $J$ so
that, for example, if $J>1$ then \nmax\ $<$ \ath\ for that channel. We can also
see here that the excitation energies compare well to the existing LS
calculations.

The lower panes show the dependence on the initial truncation of the $A$=3
sector, \ath. On the left we see that the Hamiltonian is well converged when
\ath\ = 40. On the right the excitation energy of the 3$^+$ state drops
significantly with increasing \ath. In these calculations, not only is \ath\ 
larger, but individual channels are truncated less severely as well to provide
more of the Hamiltonian for these higher $J$ states. Specifically, at \ath\ = 36
each step up in $J$ corresponded to a reduction in \nmax\ of 4. But, for the
\ath = 40 calculations this was changed to 2 \nmax\ for every
step in $J$. It is possible to push \ath\ and these channel truncations even
higher and may be needed in the future for increased accuracy.

\begin{table*}[tbh-]
\caption{\label{tab:converge} Results for binding energy in all calculations for
$^3$H and $^4$He.
All errors for SRG results are convergence margins at the quoted basis size. The
columns marked "NN+3N" show NN+NNN-induced values except for rows that include
an initial NNN. Basis sizes are \ath\ = 40 and \atw\ = 196 for all except
N$^3$LO calculations used \ath\ = 32.}
\begin{center}
\begin{tabular}{|c||c|c|c||c|c|c||c|}
\hline
\hline
 Nucleus/      & \hw\ & NN-only  		& NN+3N 			    & NN-only  	    & NN+3N		   & LS \\
 Potential     &  	 &$\lambda = 1.0$ & $\lambda = 1.0$  & $\lambda = 1.8$ & $\lambda = 1.8$ &  \\
\hline
\hline
$^3$H -- av18  &  28/52  & $-$7.487(1) &  $-$7.486(40) & $-$8.467(1) & $-$7.486(40) & $-$7.62(0)  \\	 
   cdbonn      &  28/52  & $-$7.505(1) &  $-$7.863(40) & $-$8.553(1) & $-$7.863(40) & $-$7.99(1)  \\
   n3lo        &  20  & $-$7.471(2) &  $-$7.852(5)  & $-$8.351(1) & $-$7.852(5)    & $-$7.852(5) \\				
   +NNN 	      &  20  &  ---          &  $-$8.473(5) & ---         &  $-$8.473(5) & $-$8.473(5) \\				
\hline
$^4$He -- av18 &  44  & $-$24.419(23) & $-$24.339(14) & $-$29.267(15) &  $-$23.904(25) & $-$24.23(1) \\   	   
   cdbonn      &  44  & $-$24.484(16) & $-$26.217(9)  & $-$29.739(11) &  $-$25.926(17) & $-$26.1(1)  \\		   		
   n3lo        &  28  & $-$24.284(0)  & $-$25.641(1)  & $-$28.446(1)  &  $-$25.325(1)  & $-$25.39(1) \\ 		   	
   +NNN        &  28  &  ---  		  & $-$28.661(3)  & ---           &  $-$28.464(2)  & $-$28.50(2) \\   
\hline
\end{tabular}
\end{center}
\end{table*}

Table \ref{tab:converge} gives a summary of the levels of convergence achieved
in the present calculations. These are unextrapolated results from
complete model spaces for the purposes of comparison to existing and future results and
experiment. Extrapolated results for $^6$Li will be given below. We strongly
advise the reader that the specific choice of $\lambda$ is less important than
the {\it $\lambda$ dependence} in the final results. The $\lambda$ dependence is the
indicator that many-body forces are being induced to account for the
renormalized components. Here we choose to display $\lambda$'s = 1.0 and 1.8
$\fmi$ as they reach over the range of $\lambda$ dependence in this work. For
the potentials AV18 and CD-Bonn the value for NN+NNN-induced in the table corresponds to
the unevolved minimum. The optimal frequency for the unevolved potential in these
cases is \hw\ $\simeq$ 52 MeV and these values are quoted in the table.


\section{Evolution of Many-Body Forces}
\label{sec:evolution}

\begin{figure}[thb]
\includegraphics*[width=7.7cm]{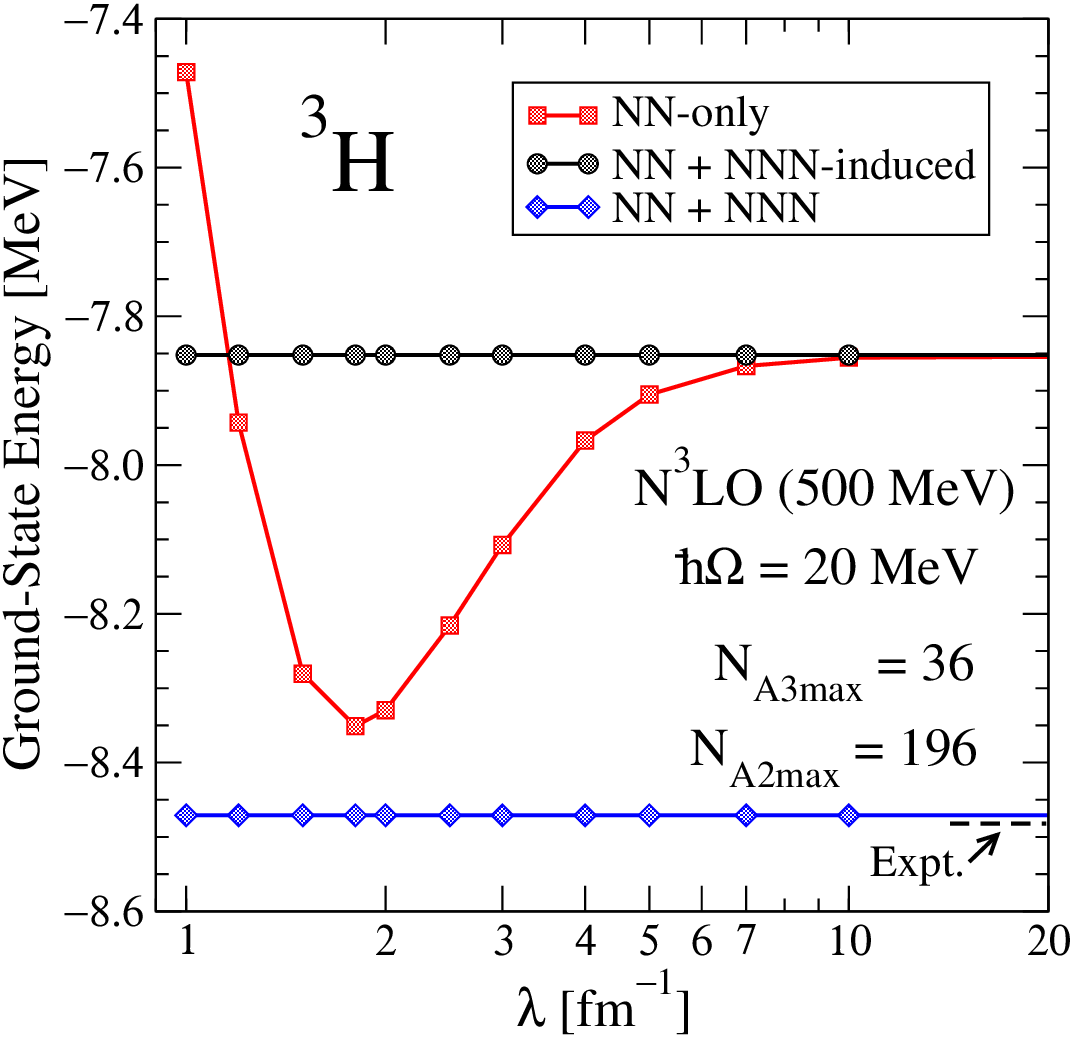}
\caption{(Color online) Ground-state energy of $^3$H as a
function of the SRG evolution parameter, $\lambda$. See
Table~\ref{tab:one} for the nomenclature of the curves. The calculations use \atw\ = 196, \ath\ = 36,
and \hw\ = 20 MeV.}
\label{fig:h3_srg}
\end{figure}


In Fig.~\ref{fig:h3_srg}, the general flow of an SRG evolved
nuclear Hamiltonian is illustrated. The ground-state energy of the triton is plotted as a
function of the flow parameter $\lambda$ from $\infty$, which is the initial (or
``bare'') interaction, toward $\lambda = 0$. We used \atw\ = 196, \ath\ = 36,
and \hw\ = 20 MeV, for which all energies are converged to better than 10\,keV.
The previous work~\cite{Jurgenson:2009qs} used a less stringent \atw\ = \ath\ =
36. However, those results are within 1 keV of the current calculations, showing
that the larger \atw\ is not critical for $^3$H.

We first consider the NN-only curve (squares). If $H_\lambda$ is evolved in only
an $A=2$ system, higher-body induced pieces are not included. The resulting
energy calculations will be only approximately unitary for $A>2$ and the
ground-state energy  will vary with $\lambda$ (squares). Keeping the induced NNN
matrix elements, by performing the $A$ = 3 evolution, yields a flat line
(circles), which confirms an exactly unitary transformation in that sector. The
line is equally flat if an initial NNN is included (diamonds). Note that the net
induced three-body is comparable to the initial NNN contribution and thus is of
natural size.

The shape of the NN-only curve, first increasing binding and then rebounding,
can be understood from the fact that early in the evolution the high-momentum
matrix elements are most affected by the transformations. These are
predominantly the short-range repulsive parts of the potential and are
transformed into strength in the induced NNN matrix elements. As the evolution
progresses to lower momentum scales the more attractive parts of the potential
are affected, causing a rebound in the NN-only result. Note that all this
information is not lost but merely reorganized into NNN terms, and when we keep
those terms we regain the unitary result. This analysis is supported by studies in
one-dimensional models that showed purely attractive initial potentials to have
monotonically decreasing binding energy~\cite{Jurgenson:2008jp,thesis}.

\begin{figure}[htb]
\includegraphics*[width=7.7cm]{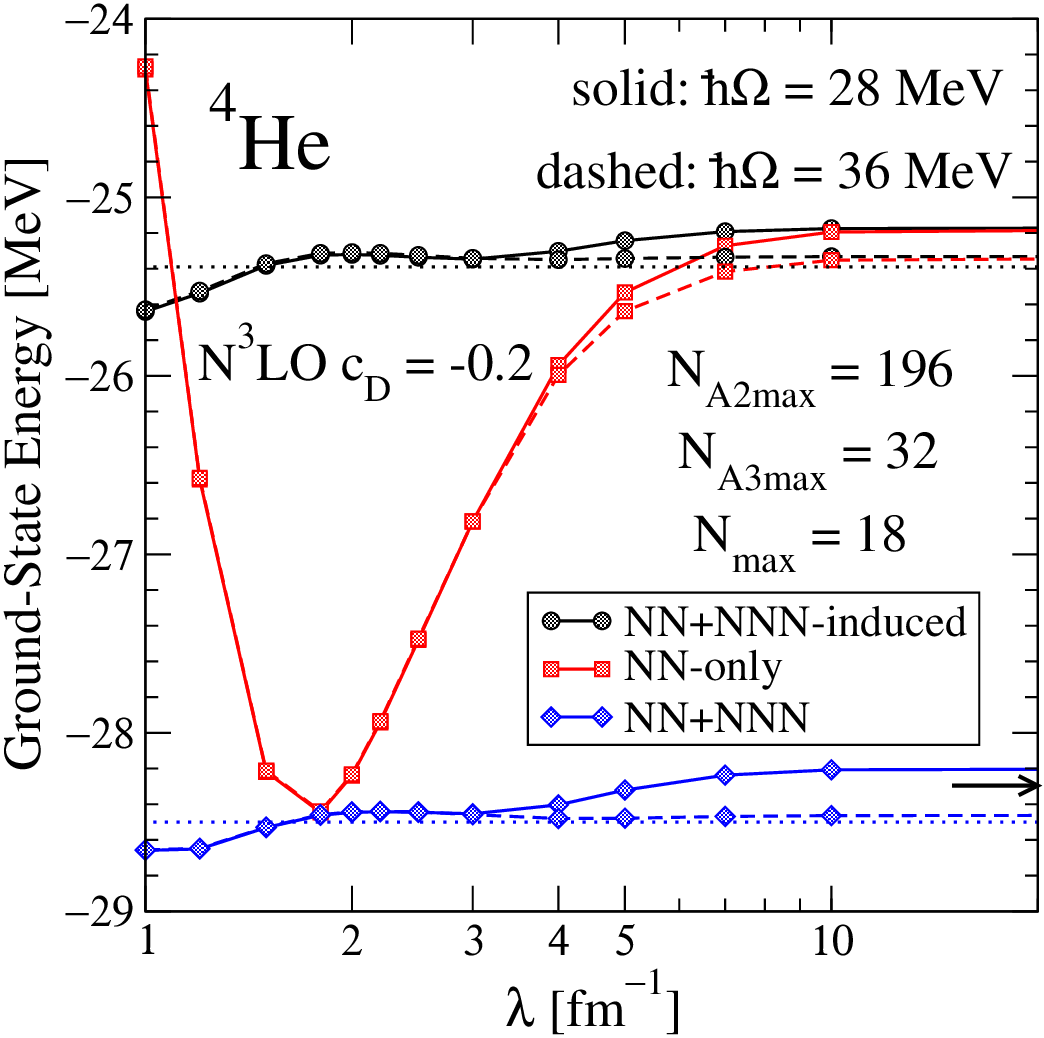}
\caption{(Color online) Ground-state energy of $^4$He as a function of the SRG
evolution parameter, $\lambda$. Note the comparison of two values for \hw\
(solid and dashed) and the best LS results (dotted). Here, \atw\ = 196 and \ath\
= 32 compared to \atw\ = \ath\ = 28 for the prior work~\cite{Jurgenson:2009qs}.
The results here are converged to within 10 keV of those from
Ref.~\cite{Jurgenson:2009qs}. See Table~\ref{tab:one} for the nomenclature of
the curves. Note the comparison of two values for \hw\ (solid and dashed) and
the best LS results (dotted). The thick arrow indicates the experimental value.}
\label{fig:he4_srg}
\end{figure}


In Fig.~\ref{fig:he4_srg}, we examine the SRG evolution in $\lambda$ for $^4$He
using a chiral N$^3$LO potential~\cite{N3LO} with \hw\ = 36 and 28 MeV, the
dashed and solid sets respectively. The $\la V_\lambda^{(2)}\ra$ and $\la
V_\lambda^{(3)}\ra$ matrix elements were evolved with basis sizes \atw\ = 196
and \ath\ = 32 and then truncated to \nmax\ = 18 at each $\lambda$ to
diagonalize $^4$He. Again, the higher \atw\ has little impact on the final
results for this nucleus. The NN-only curve has the characteristic shape
discussed above. When the induced NNN is included, the $\lambda$ dependence is
significantly reduced. The pattern only depends slightly on the inclusion of
initial NNN interaction. In both cases the dotted line represents the converged
value for the initial Hamiltonian using a Lee-Suzuki based procedure. The
residual $\lambda$ dependence is due to missing induced four-body forces.

At large $\lambda$, the discrepancy with the dotted line is due to a lack of
convergence for unevolved potentials at \nmax\ = 18, but at $\lambda < 3 \fmi$
SRG decoupling takes over and the discrepancy is due to short-range induced
four-body forces. This transition is emphasized by showing the calculation at
two different values of \hw. The point in $\lambda$ where they meet is an
indicator of the momentum scale at which the evolving Hamiltonian is converged
with \nmax\ = 18. All the information included in the initial Hamiltonian at
\ath\ has been transformed into the smaller basis defined by \nmax. Any residual
difference between values of \hw\  (invisible on this scale) indicate the level
of convergence with respect to the included $\la V_\lambda^{(3)}\ra$ matrix
elements defined by \ath.

In the three-body-inclusive curves the discrepancy due to induced four-body
forces is about 50\,keV net at $\lambda = 2\,\mbox{fm}^{-1}$.  This is small
compared to the rough estimate in Ref.~\cite{Rozpedzik:2006yi} that the
contribution from the long-ranged part of the N$^3$LO four-nucleon force to
$^4$He binding is of order a few hundred keV. If needed, we could evolve 4-body
matrix elements in $A=4$ and will do so when nuclear structure codes can
accommodate them.

\begin{figure}[htb]
\includegraphics*[width=7.7cm]{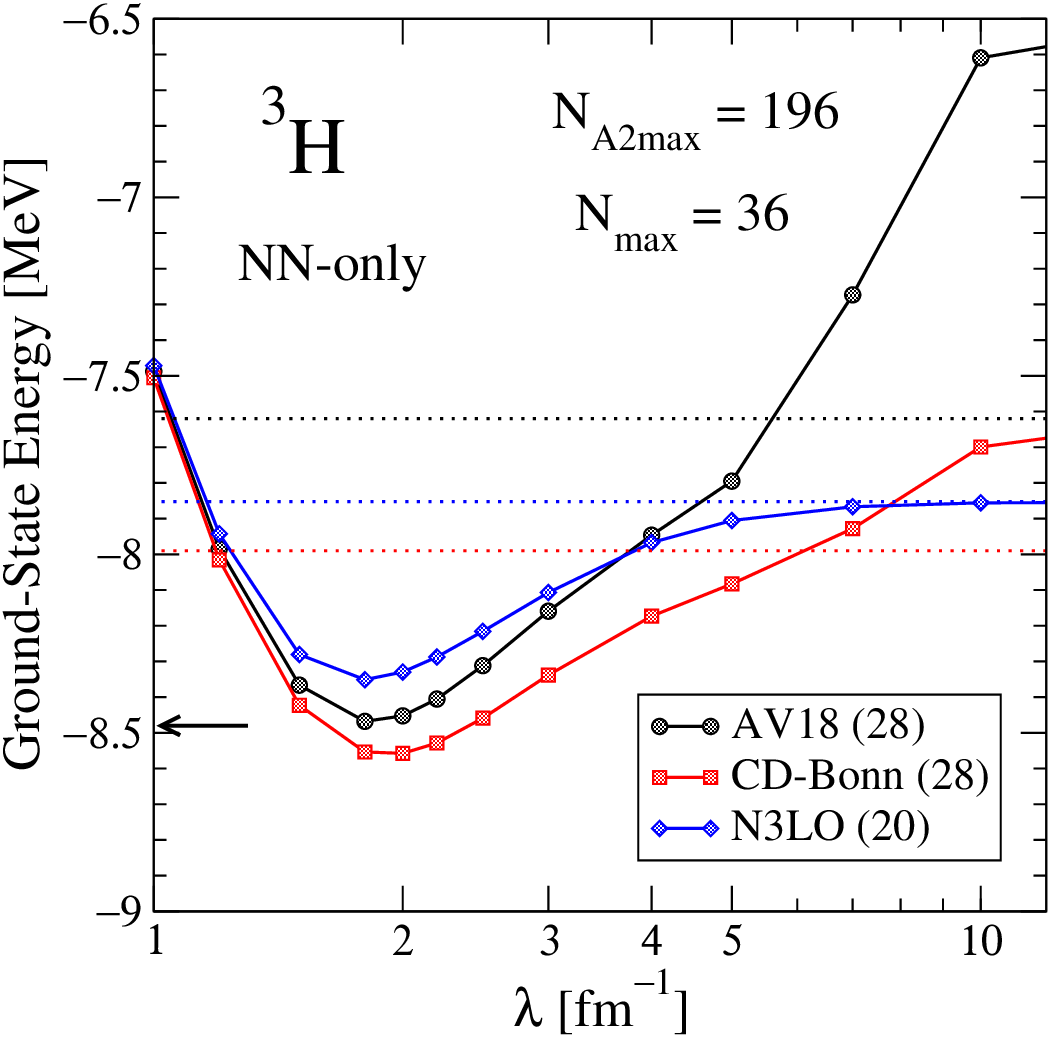}
\includegraphics*[width=7.7cm]{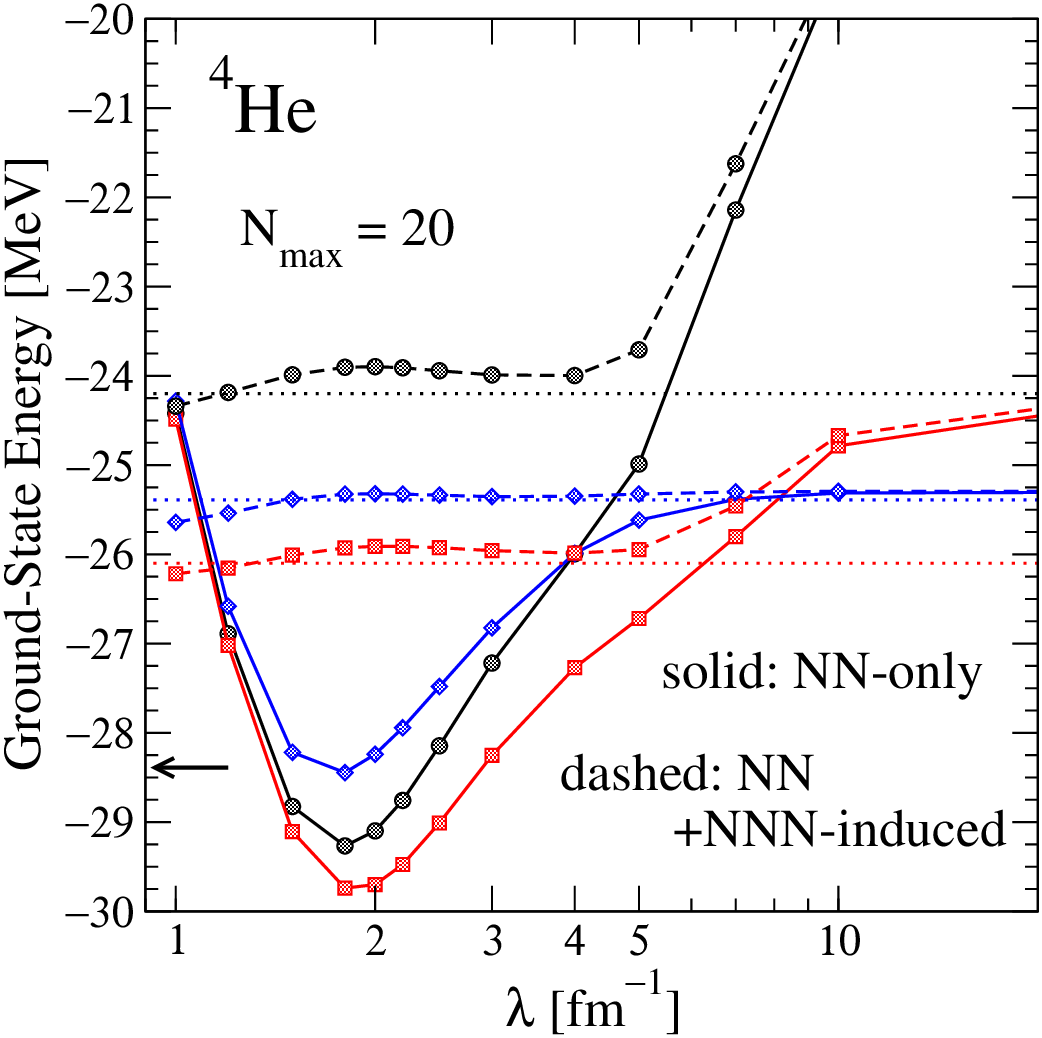}
\caption{(Color online) Ground-state energy of $^3$H and $^4$He as a function of
$\lambda$, starting from Hamiltonians based on CD-Bonn~\cite{pot_cdbonn},
Argonne $V_{18}$~\cite{pot_av18}, and N$^3$LO~\cite{N3LO}. See
Table~\ref{tab:one} for the nomenclature of the curves. Here we used \hw\ =
28/28/20 and 44/44/28 for $^3$H and $^4$He respectively. The dotted lines show
the converged LS results for each potential. For $^3$H these values are AV18 =
$-7.62$, CD-Bonn = $-7.99$, and N$^3$LO = $-7.85$.}
\label{fig:He4_av18_cdbonn}
\end{figure}


Figure \ref{fig:He4_av18_cdbonn} compares the flow of $^3$H and $^4$He binding
energies for several initial Hamiltonians, AV18~\cite{pot_av18},
CD-Bonn~\cite{pot_cdbonn}, and N$^3$LO~\cite{N3LO}. For $^3$H we have used
harmonic oscillator frequencies \hw\ = 28, 28, and 20 MeV respectively. For
$^4$He these optimal frequencies are  \hw\ = 44, 44, and 28 MeV respectively.
The general shape of all the NN-only curves is quite similar here with a initial
dip in binding and then a turn over at $\lambda = 1.8 \fmi$. Evolution to low
$\lambda$ ($<$ 2.0 $\fmi$) of different initial Hamiltonians at $A=2$ produce a
very similar evolved form~\cite{Bogner:2009bt}. Previously this had only
been observed at the level of comparing selected two-body matrix elements. Here
we show that observables at $A>2$ also exhibit this behavior. The NN-only points
for all three initial Hamiltonians converge as $\lambda$ decreases past $1.8$. 

Note that the values for the unevolved potentials AV18 and CD-Bonn do not
approach the Lee-Suzuki results because these potentials require a larger $UV$
cutoff and therefore a larger \hw\ ($\simeq 52$) or \nmax\ for convergence. The
converged values for the bare potentials are quoted in Table~\ref{tab:converge}
and would be the equivalent of the unitary line shown in Fig.~\ref{fig:h3_srg}.

Of course, we should not expect the NN+NNN-induced calculations to produce
identical results at small $\lambda$ because they are not equivalent
Hamiltonians at the $A=3$ level. However their very similar shape indicates a
specific scale dependence of three- and four-body forces generated during
evolution. This is reminiscent of evolution of the chiral interaction in
Fig.~\ref{fig:he4_srg} with and without initial NNN, where the shape is similar
but shifted by the initial difference.  This is a promising indication of a
universality phenomenon. A full test of this idea will require coding analogous
three-body interactions for the other initial potentials (i.e., IL-IX for AV18)
and evolving other initial NN interactions.

\begin{figure}[htb]
\includegraphics*[width=8cm]{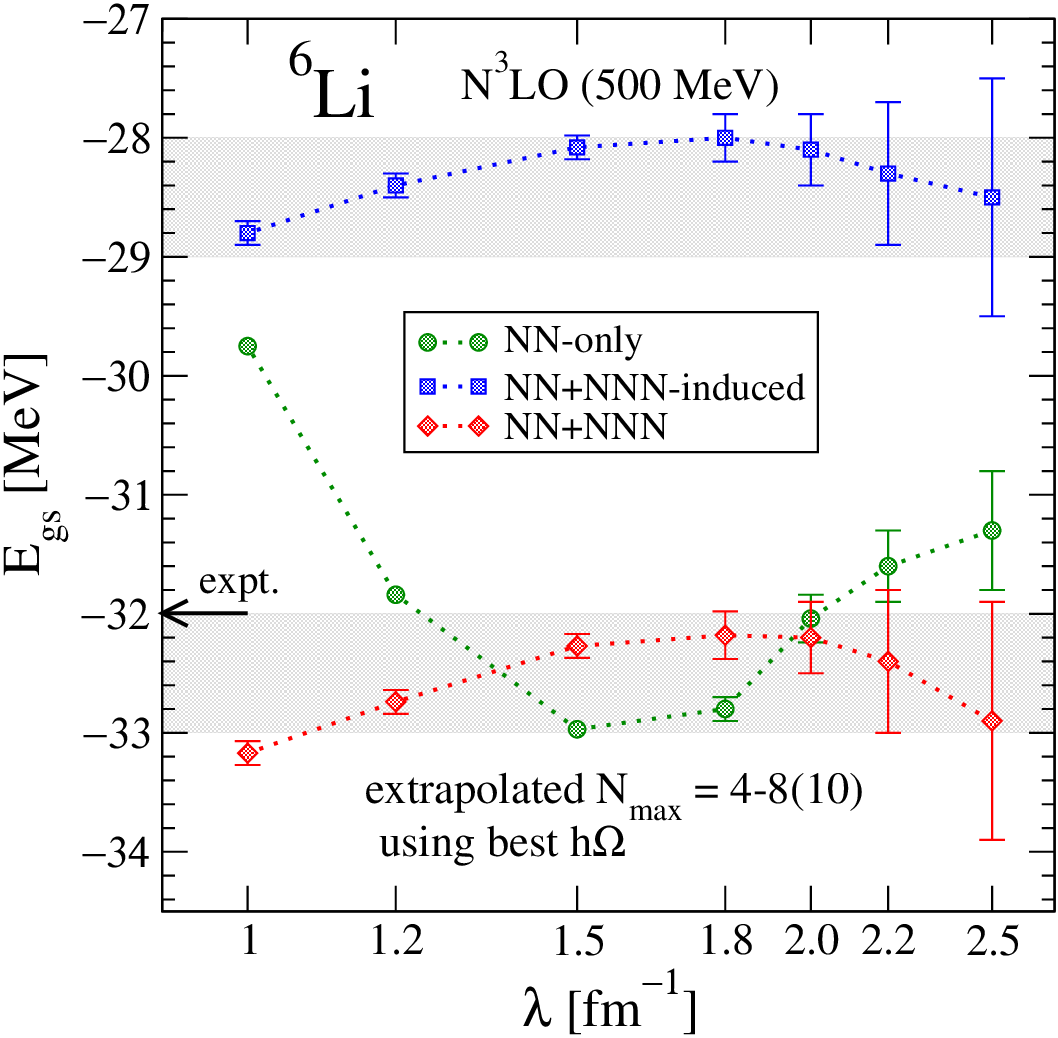}
\caption{(Color online) Extrapolated ground-state energy of $^6$Li as a function
of the SRG evolution parameter, $\lambda$. Error bars are based on variation of
the minimum in \hw\ as discussed in the text. The truncations used were \atw\ =
300 and \ath\ = 40. }
\label{fig:Li6_srg} 
\end{figure}


Figure \ref{fig:Li6_srg} shows extrapolated ground-state energies for $^6$Li at
different values of $\lambda$. We used truncations \atw\ = 300 and \ath\ = 40
and performed the final diagonalization up to \nmax\  = 8. The  gray bands
represent the best Lee-Suzuki results available for NN-only and NN+NNN initial
interactions, with error estimates. The analogous truncations for these
calculations were \atw\ = 400 and \ath\ = 40 with $^6$Li being calculated up to
\nmax\ = 14 for NN-only. The results are $28.5 \pm 0.5$ MeV without and $32.5 \pm 0.5$ MeV
with initial three-body forces.

The $\lambda$ dependence is shown for the lower values where the result is
near convergence. The results plotted here are obtained from the extrapolation
procedure previously described. This procedure accounts for the variation, with
$\lambda$, of the minimum \hw\ and extrapolates in \nmax\ the converged binding energy.
The error bars are dependent on the range in \hw\ for which we have results. For
any given $\lambda$ the error bars will be larger if the optimal \hw\ is not
present in the data set used for extrapolation. In fact the extrapolation tends
to predict too low (more negative) as measured by the predictions of \nmax\ = 10
points in the NN-only curve. This feature has also been confirmed as we
systematically added to the data set for the $\lambda$'s with the largest error
bars; The extrapolated points rose with better values of \hw, flattening the
curve and reducing the apparent $\lambda$ dependence.

The hierarchy of induced many-body forces can be assessed in
fig.~\ref{fig:Li6_srg} by comparing the spread in the NN-only curve to that of
the NN+NNN curves. To do so, note that the NN-only curve must coincide with the
LS result at $\lambda = \infty$. The spread in $\lambda$ has been reduced from 4
MeV to $< 1$ MeV. The majority of induced many-body forces missing from the
NN-only curve is due to three-body forces subsequently included in the other
curves.  Notice that the shape of the evolution curve is very similar to those
of $^4$He from any of the initial potentials used --- a gentle rise to 1.8 and a
slightly steeper slope down through 1.0. We interpret this to indicate that the
majority of many-body ($A\geq4$) forces induced are for $A=4$ and that 5- and
6-body forces are significantly smaller. This is consistent with the expected
hierarchical flow of induced many-body forces~\cite{Jurgenson:2008jp}.
The spread here is roughly one MeV, compared to the 30-60 keV found in $^4$He
calculations. 

Our NN-only curve is almost identical in shape to previous momentum space
studies, despite the difference in initial Hamiltonians used. The previous study
used only the neutron-proton interaction for all NN interactions while we have
used the full  isospin-breaking potential. The results are systematically
shifted up (in the previous study $^6$Li was overbound by $\simeq 1$MeV) in
relation to the NNN-inclusive curves. The error bars for the same \nmax\  are
roughly the same as the previous work, confirming that NN-only calculations are
good predictors of appropriate \hw\ values.

\begin{table*}[tbh-]
\caption{\label{tab:Li6} Results for extrapolated binding energy of $^6$Li at
various values of $\lambda$ and includes error bars from the extrapolation.  The
analogous results from Lee-Suzuki calculations are $-28.5 \pm 0.5$ MeV for
NN-only and NN+NNN-induced, and $-32.5 \pm 0.5$ MeV for NN+NNN. The experimental
value is $-31.99$ MeV. The \hw\ and range in \nmax\ used for each extrapolation
is also quoted. } 
\begin{center}
\begin{tabular}{|c||c|c|c|c|}
\hline
\hline
 $\lambda$ & best \hw\  & NN-only & NN+NNN-induced & NN+NNN  \\
 			   &           & (\nmax\ 4-10) &  (\nmax\ 4-8) &  (\nmax\ 4-8)  \\
\hline
  2.5  & 24 &  $-$31.3 $\pm$ 0.5   & $-$28.5 $\pm$ 1.0 &	-32.9 $\pm$ 1.0  \\ 	 
  2.2  & 20\footnote{This one point has a different optimal \hw\ for NN-only at
 \hw\ = 24.} &  $-$31.6 $\pm$ 0.3   & $-$28.3 $\pm$ 0.6 &	-32.4 $\pm$ 0.6  \\ 	  
  2.0  & 20 &  $-$32.0 $\pm$ 0.2   & $-$28.1 $\pm$ 0.3 &	-32.2 $\pm$ 0.3  \\ 	  
  1.8  & 16 &  $-$32.8 $\pm$ 0.1   & $-$28.0 $\pm$ 0.2 &	-32.2 $\pm$ 0.2  \\ 	
  1.5  & 16 &  $-$33.00 $\pm$ 0.05 & $-$28.1 $\pm$ 0.1 &	-32.3 $\pm$ 0.1  \\  
  1.2  & 16 &  $-$31.85 $\pm$ 0.05 & $-$28.4 $\pm$ 0.1 &	-32.75 $\pm$ 0.1 \\ 	 
  1.0  & 16 &  $-$29.75 $\pm$ 0.02 & $-$28.8 $\pm$ 0.1 &	-33.2 $\pm$ 0.1  \\ 	 
\hline
\end{tabular}
\end{center}
\end{table*}


Note that the many-body forces do not explode as previously feared and that
evolution to lower $\lambda$ may not be unreasonable. While previous NN-only
studies showed induced three-nucleon forces growing uncontrolled below
$\lambda=1.5$, we see that inclusion of these matrix elements produces a more
gentle $\lambda$ dependence. So, evolving to lower $\lambda$ to improve
convergence may be useful in future calculations. Recent results~\cite{Roth_pc}
suggest that, for the choice of SRG generator in Eq.~\eqref{eq:flow}, the
many-body forces may grow with $A$, so monitoring in the rest of the p-shell
(with adequate codes) is vital. Alternative choices for the SRG generator and
sophisticated extrapolation techniques may play a central role.


\section{Hierarchy}
\label{sec:hierarchy}

\begin{figure}[thb]
\includegraphics*[width=7.7cm]{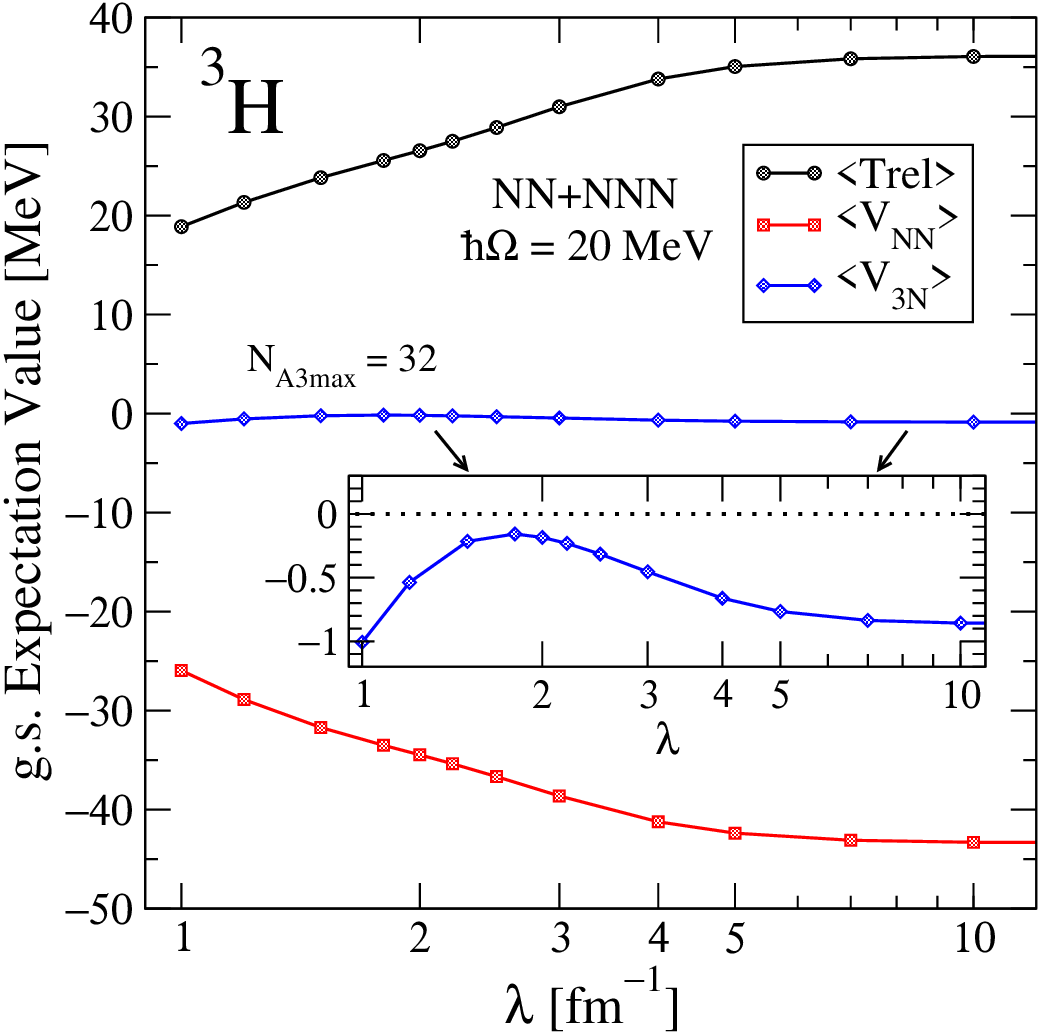}
\includegraphics*[width=7.7cm]{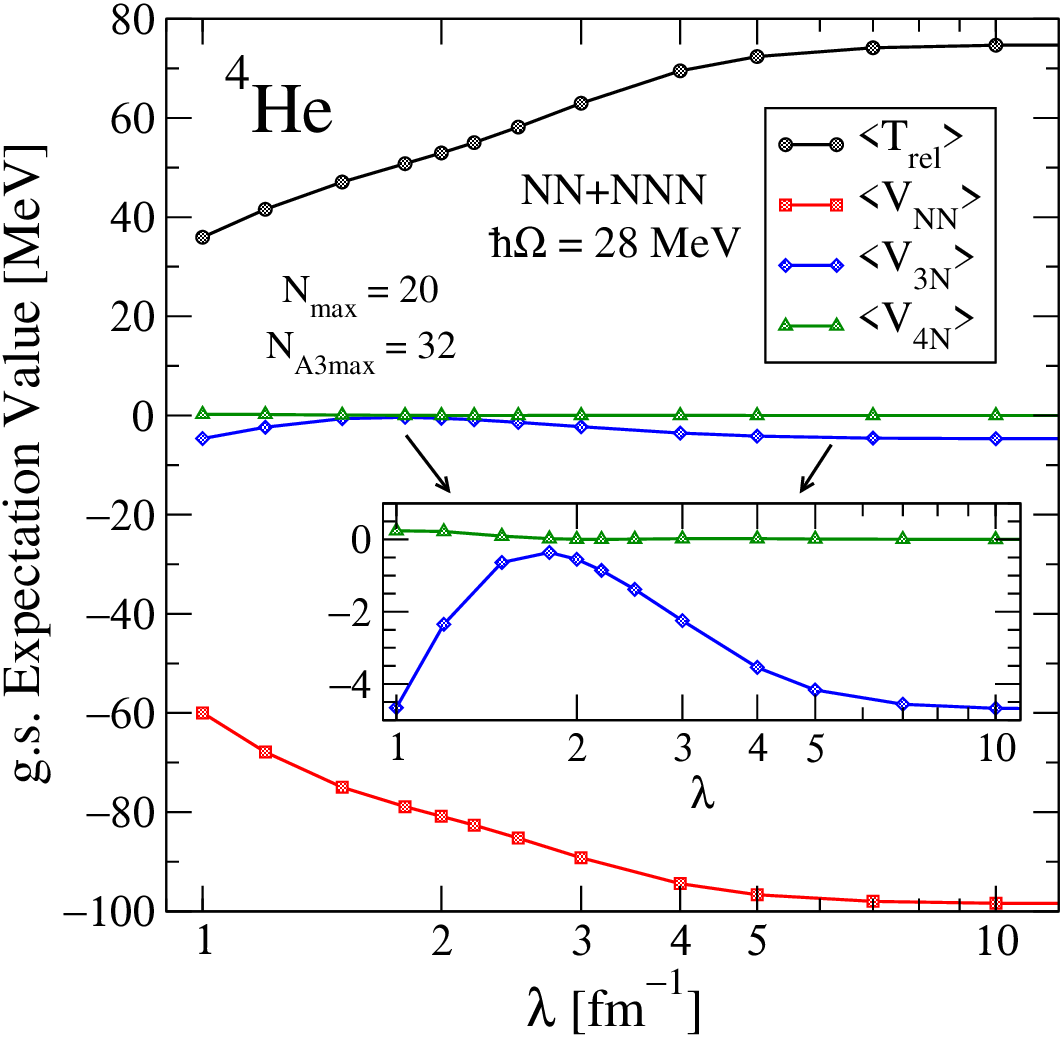}
\caption{(Color online) Expectation value of the components of the nuclear
interaction in the $^3$H and $^4$He ground states as a function of the SRG
parameter $\lambda$. The initial interaction was N$^3$LO NN~\cite{N3LO} and N$^2$LO NNN at
\hw\ = 20 and 28 MeV  (left and right) and \ath\ = 32 with the $^4$He final
truncation at \nmax\ = 18. The insets show expanded details for the 3- and 4-body forces.}
\label{fig:V_exp_val}
\end{figure}

In order to more fully understand the SRG  evolution we can trace the individual
parts of the Hamiltonian. Figure \ref{fig:V_exp_val} shows the $^3$H and $^4$He
ground-state expectation values of individual components of the evolving
Hamiltonian as a function of $\lambda$. The insets show an increased scale for
closer inspection of 3- and 4-body expectation values. Here a hierarchy of
induced many-body forces is evident. The magnitude of variation in $\lambda$ for each curve
differs by approximately an order of magnitude. Cancellations between $T_{rel}$
and $V_{NN}$ are reduced significantly over the course of the evolution. The
strength of matrix elements at large momenta is being  reorganized into (shifted
to) lower momentum matrix elements. Hence the absolute values of the 
expectation values $\la T_{rel} \ra$ and $\la V_{NN} \ra$ are reduced. 

 Also, note the correspondence between the $\la V_{3N} \ra$ curve and NN-only
evolution curves such as in Figs.~\ref{fig:h3_srg} and \ref{fig:he4_srg}. The
size of the three-body force reaches a minimum corresponding to the point
($\lambda \simeq 1.8 \fmi$) of maximum binding achieved by the NN-only
calculations. This is simply the explicit plotting of the many-body forces that
is implied by the approximately unitary curves shown in
Section~\ref{sec:evolution}. In this case, the induced three-body forces
effectively cancel out the initial three-body terms; the expectation value, $\la
V_{3N} \ra$, drops almost to zero.

To make a connection between the individual terms in the three-body interaction
evolution and the running of the ground-state energy, we need the
evolution equations for the \emph{expectation value} of $V^{(3)}_s$ in the
ground state. Denoting the ground-state wave function for the $A$-particle
system by $|\psi^A_s\ra$, it evolves according to (it is convenient here to use
the flow parameter $s = 1/\lambda^4$)
\beqn
  |\psi^A_s\ra = U_s |\psi^A_{s=0}\ra \;,
  \quad \mbox \quad \frac{d}{ds} |\psi^A_s\ra = \eta_s |\psi^A_s\ra 
  \;,
\eeqn
where $U_s$ is the SRG unitary transformation at $s$ and
\beqn
  \eta_s = \frac{dU_s}{ds} U^\dagger_s = - \eta^\dagger_s \;.
\eeqn  
Then the matrix element of an operator $O_s$ evolves according to
\beqn
  \frac{d}{ds} \la \psi^A_s | O_s | \psi^A_s \ra
  = \la \psi^A_s | \frac{d O_s}{ds} - [\eta_s,O_s] | \psi^A_s \ra
  \;.
  \label{eq:Oevolve}
\eeqn
If the operator $O_s$ is transformed as $O_s = U_s O_{s=0}
U_s^\dagger$, then the matrix element on the right-hand-side of
Eq.~\eqref{eq:Oevolve} vanishes, as when $O_s = H_s$.

However, if we wish to see how one part of $H_s$
evolves, such as the expectation value of $V^{(3)}$, we obtain
\beqn
  \frac{d}{ds}\la \psi^A_s |V_s^{(3)}|\psi^A_s\ra = 
  \la \psi^A_s|\frac{dV^{(3)}_s}{ds} - [\eta_s,V_s^{(3)}]|\psi^A_s\ra
  \;,
\label{eq:dds_vev}
\eeqn
which does not give zero in general because $V_s^{(3)} \neq
U_sV_{s=0}^{(3)}U_s^{\dagger}$.  In the two-particle case, the analog of
Eq.~\eqref{eq:dds_vev} gives $d\la V^{(2)} \ra/ds = \la [\eta_s,T_{\rm rel}]
\ra$. In the three-particle case, we can expand Eq.~\eqref{eq:dds_vev} as
\bea
	\frac{d}{ds}\la \psi^A_s |V^{(3)}_s|\psi^A_s\ra &=& \la \psi^A_s|[\eta_s,H_s]_3 -
		[\eta_s,V^{(3)}_s]|\psi^A_s\ra  \nonumber \\
	&=& \la \psi^A_s | [\vbtr,T_{\rm rel}] + [\vbt,V^{(2)}_s]_c + [\vbt,V^{(3)}_s] + [\vbtr,V^{(2)}_s] +
		[\vbtr,V^{(3)}_s] \nonumber \\
	&& \qquad - [\vbt,V^{(3)}_s] - [\vbtr,V^{(3)}_s]  | \psi^A_s \ra  \nonumber \\
	&=& \la \psi^A_s |  [\vbtr,H_s] + [\vbt,V^{(2)}_s]_c - [\vbtr,V^{(3)}_s]| \psi^A_s \ra \nonumber \\
	&=& \la \psi^A_s | [\vbt,V^{(2)}_s]_c - [\vbtr,V^{(3)}_s]| \psi^A_s \ra 
	\;,
	\label{eq:A3vevs}
\eea
where $\vbt$ and $\vbtr$ are the commutators $\vbt \equiv [T_{\rm rel},V^{(2)}_s]$
and $\vbtr \equiv [T_{\rm rel},V^{(3)}_s]$. In the third line, the expectation value
of the commutator, $[\vbtr,H_s]$, vanishes identically.

The subscript ``$c$'' in the first term indicates that only connected parts of
this commutator have been kept, and refers to a diagrammatic formalism developed
in Ref.~\cite{Bogner:2007qb}. Computing $[\vbt,V^{(2)}]$ in the three-particle
space involves all nucleons democratically. However, commutators which leave one
nucleon as a spectator cancel out in the $A$ = 2 sector. So, we must compute
$[\vbt,V^{(2)}]$ for the $A=2$ sector and embed it in $A$ = 3 so we can
isolate the piece that affects the evolution of $V^{(3)}$. In general, this
subtraction is required at all sectors in $A$, and the ``$c$'' here indicates that this procedure
has been done.

\begin{figure}[thb]
\includegraphics*[width=7.7cm]{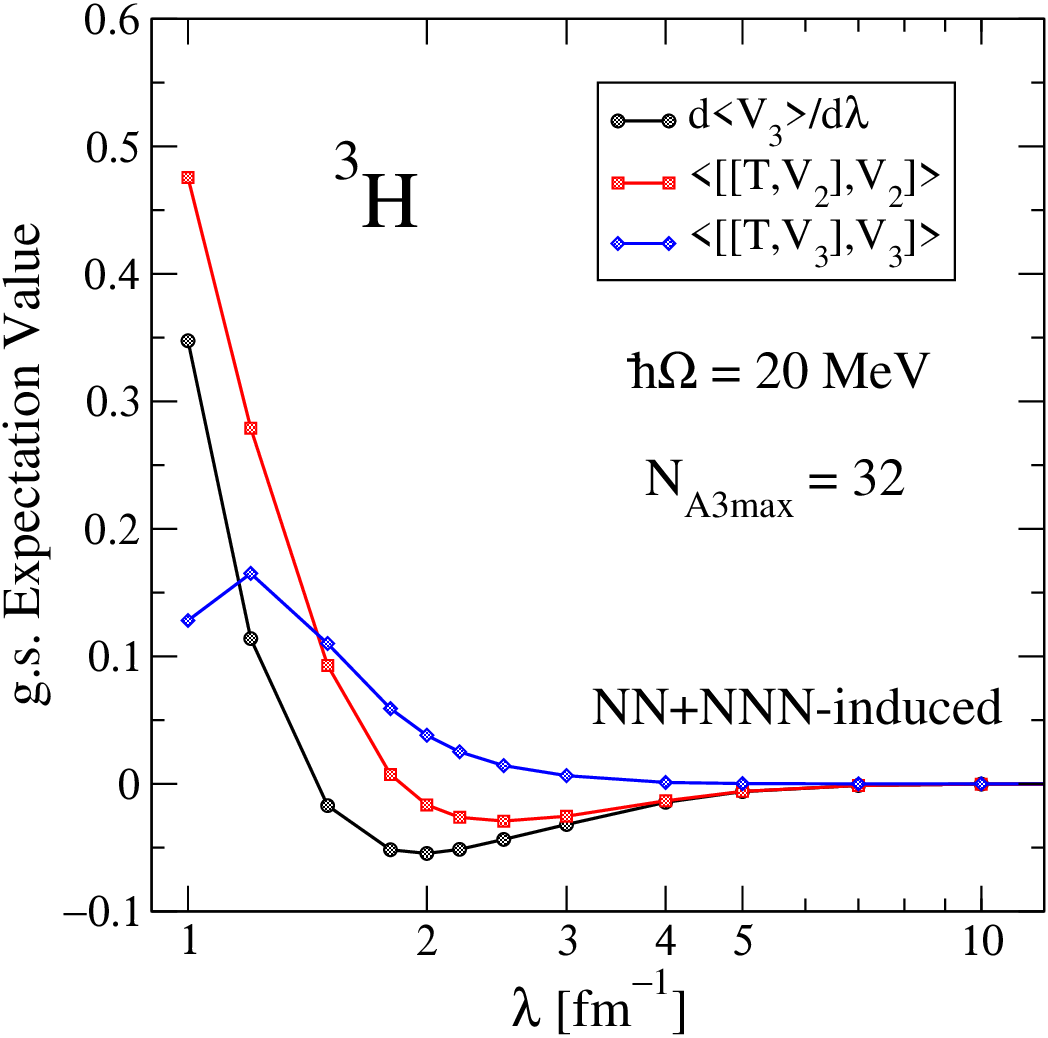}
\includegraphics*[width=7.7cm]{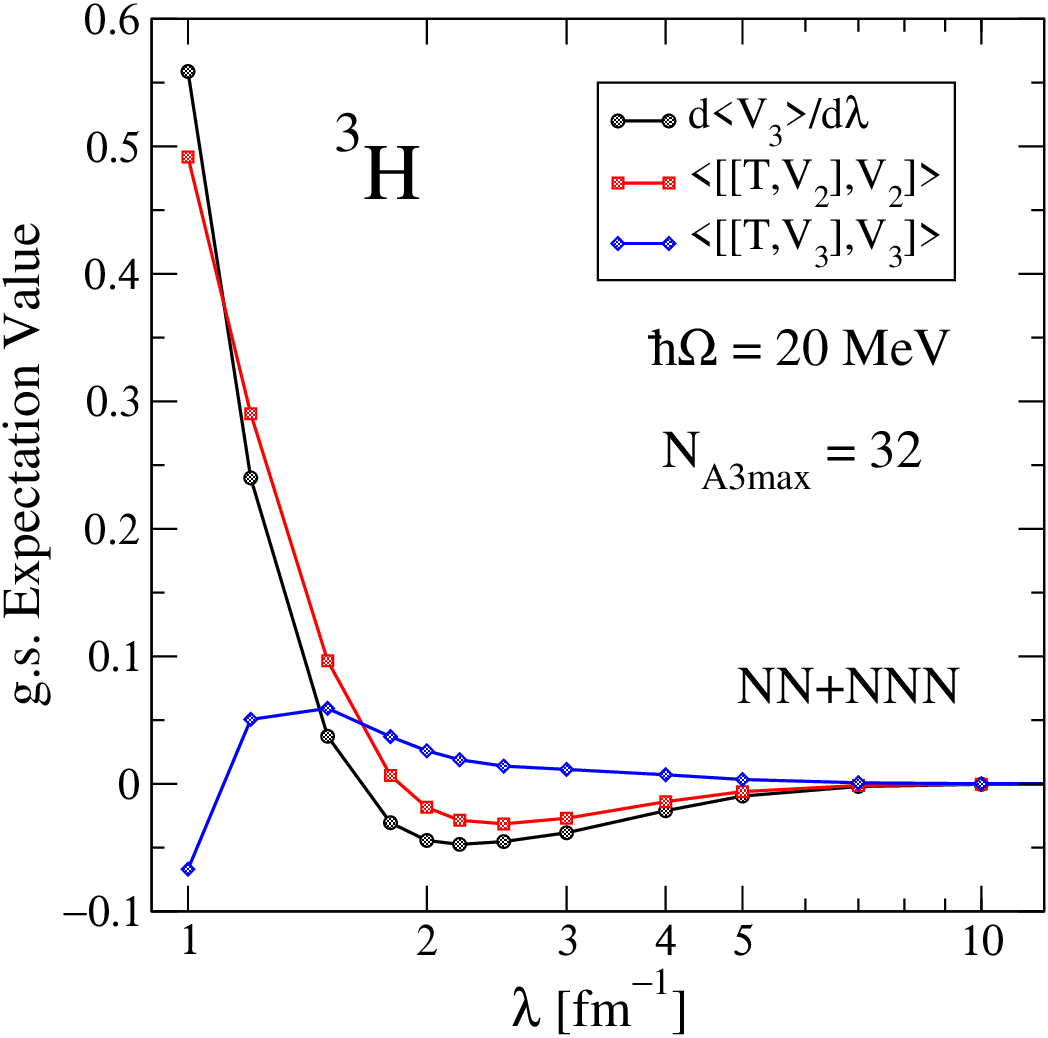}
\caption{(Color online) Contributions from individual terms to $d \la
V^{(3)}_\lambda\ra / d \lambda$, the flow of the triton ground-state expectation value of
the three-body potential. On the left (right) the calculation is done without
(with) an initial three-body interaction.}
\label{fig:srg_VEV3}
\end{figure}

In Fig.~\ref{fig:srg_VEV3} we show the ground-state expectation values of the
terms in Eq.~\eqref{eq:A3vevs} for the triton. The left panel shows the calculations with just
induced NNN interactions and the right panel with an initial three-body force.
It is most useful for our analysis to convert from derivatives with respect to
$s$ to derivatives with respect to $\lambda$ using $\frac{d}{ds} =
-\frac{\lambda^5}{4}\frac{d}{d\lambda}$. The dominant contribution to the
evolution of the three-body potential matrix element is the two-body connected
part,  $[\vbt,V^{(2)}_s]_c$. This dominance is stronger here than seen in a one
dimensional analog~\cite{Jurgenson:2008jp} perhaps due to a stronger initial
hierarchy in the EFT compared to the initial conditions chosen in one-dimension.
Again, the evolution of three-body matrix elements depends on the interplay between
long- and short-range, attractive and repulsive parts, which lead to
scale-dependent inflection points and slopes.

We can repeat the above analysis for $A = 4$ and obtain
\bea
\frac{d}{ds}\la
\psi^{(4)}_s|V_s^{(4)}|\psi^{(4)}_s\ra  
&=& \la \psi^{(4)}_s|[\overline{V}^{(2)}_s,V^{(3)}_s]_c 
+ [\overline{V}^{(3)}_s,V^{(2)}_s]_c  \nonumber \\
&& \quad + [\overline{V}^{(3)}_s,V^{(3)}_s]_c 
- [\overline{V}^{(4)}_s,V^{(4)}_s]| \psi^{(4)}_s \ra \;,
\label{eq:A4vevs}
\eea
where we find no fully connected terms with only two-body forces. Again,
disconnected terms involving two and three body potentials cancel out in the
lower sectors. The leading terms are commutators with one $V^{(2)}_s$ and one
$V^{(3)}_s$, followed by connected terms quadratic in $V^{(3)}_s$ and one term
quadratic in $V^{(4)}_s$. All terms are small and additional cancellations among
them may further suppress the four-body contribution. Thus, the initial hierarchy of
many-body forces suggests that induced four-body (and higher-body) forces will be
small.

\begin{figure}[thb]
\includegraphics*[width=7.7cm]{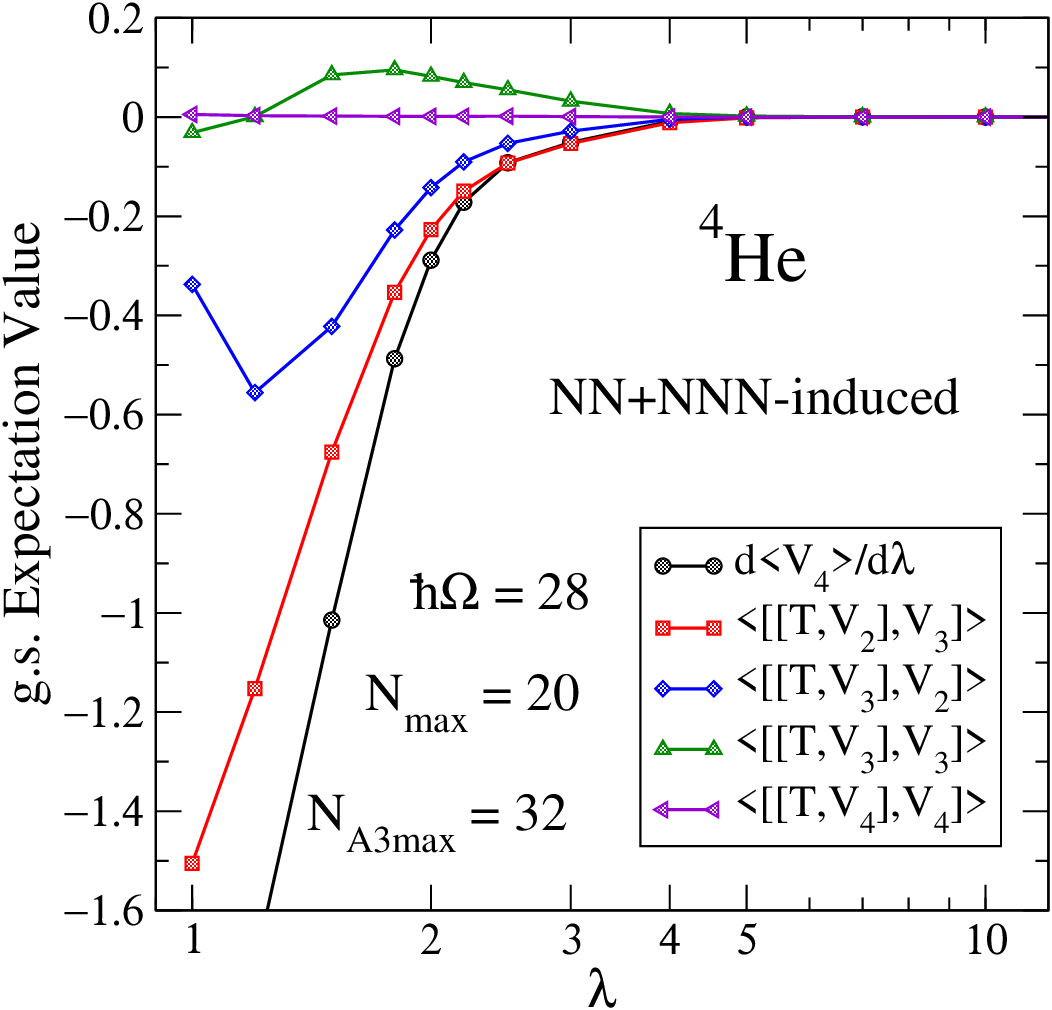}
\includegraphics*[width=7.7cm]{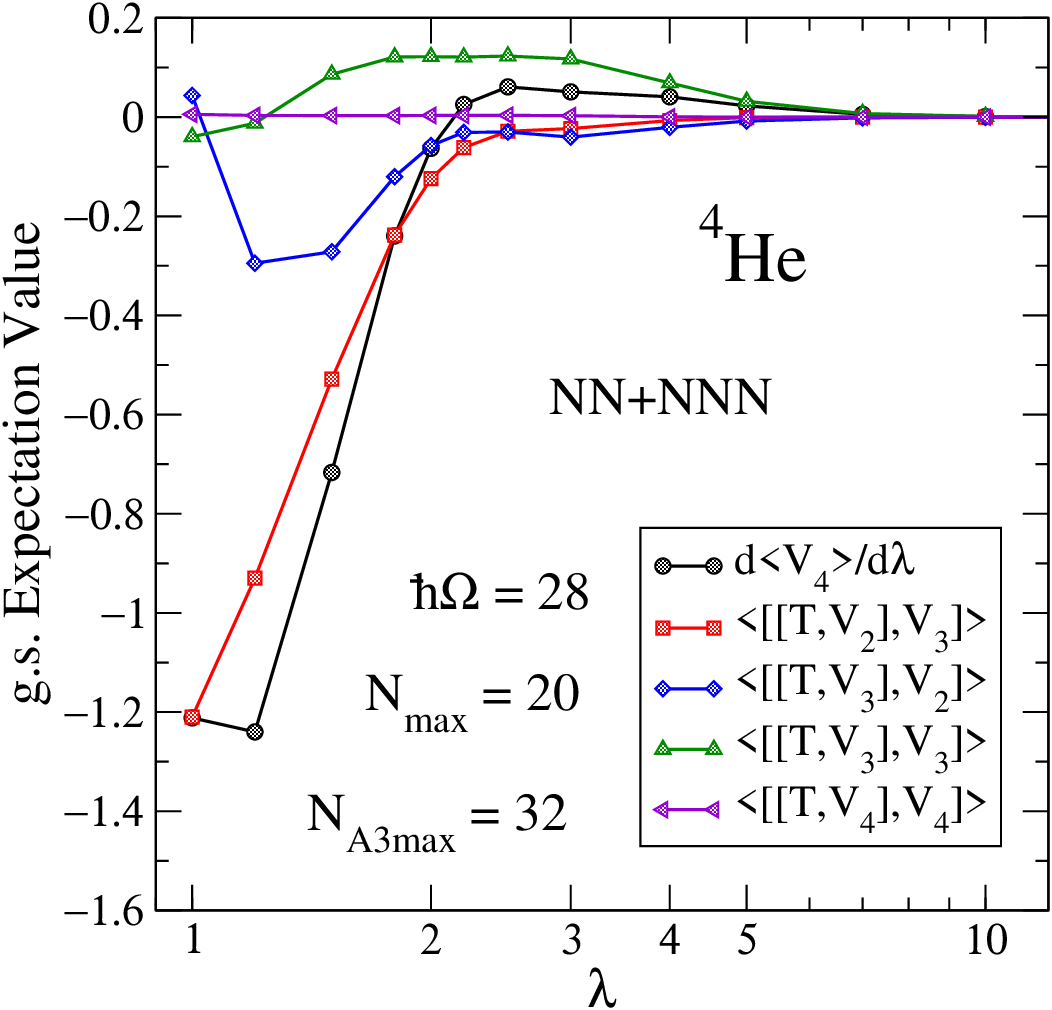}
\caption{(Color online) Contributions of the terms in Eq.~\ref{eq:A4vevs} to $d
\la V^{(4)}_\lambda\ra / d \lambda$, the evolving $^4$He ground-state expectation value
of the four-body force. On the left (right) the calculation is done without
(with) an initial three-body interaction.}
\label{fig:srg_VEV4}
\end{figure}

In Fig.~\ref{fig:srg_VEV4} we plot these contributions to the evolution of the
four-body expectation value. On the left panel is shown the calculations with
just induced NNN interactions and the right panel includes initial three-body
forces. Again it is more useful to convert the derivatives in $s$ to derivatives
in $\lambda$. The interplay of contributions is much more complicated than for
$A$ = 3. We can see cancellations between one commutator involving  $V^{(2)}_s$
and $V^{(3)}_s$ (blue diamonds), and the term quadratic in $V^{(3)}_s$ (green
triangles). This is in slight contrast to the analogous case in one dimension
where all four terms were involved in less straightforward cancellations. No
terms quadratic in $V^{(2)}_s$ appear because no connected diagrams can be
constructed for the four-particle evolution. The total derivative of $V^{(4)}_s$
is small until below $\lambda = 2$. Again the dominant contribution to the flow
is the lowest order commutator, and no feedback in $V^{(4)}_s$ is present.

There is room here for dependence of the induced many-body forces on the
strength of the initial three-body potential and figure~\ref{fig:srg_VEV4}
supports this as far as $A=4$. Forthcoming results~\cite{Roth_pc} provide
evidence of such dependence increasing with $A$. Other forms of SRG, such as one
with the replacement $T_{rel} \rightarrow T_{rel}+V_{2\pi}$, may be useful in
controlling the renormalization of the long range parts of the initial
potential.

We also note that any complete analysis of the growth of induced many-body
forces must involve converged or extrapolated results at the optimal \hw\ for
each $\lambda$. The analysis tool shown here is only meaningful when viewed at a
single \hw\ over the course of evolution in $\lambda$ and direct comparison to
plots of the type shown in figure~\ref{fig:Li6_srg} is difficult.


\section{Observables}
\label{sec:radius}


While accurate reproduction of nuclear binding energies is the first step
in nuclear structure calculations other observables can offer additional information
about the effects of renormalizing high-energy degrees of freedom, short-range
correlations, and other details of a properly fit initial Hamiltonian. While we 
know that the harmonic oscillator basis is not an ideal environment for certain
long-ranged observables, such as the rms radius, we have existing Lee-Suzuki
renormalized benchmarks with which to compare. And electromagnetic transitions,
such as B(E2)'s and B(M1)'s are notoriously difficult, both to calculate and to
measure, making this an important area of prediction for theory. All such
observables are an important next test in understanding the quality of the
many-body wavefunctions resulting from SRG-evolved interactions. Here we present
a small sampling of results, focusing on convergence patterns.

Here we are plotting the unevolved operator expectation value in the evolved
wavefunction. This is a reasonable way to visualize the effect of evolution on
the structure with respect to particular operator. However, consistent
renormalization of the operators themselves is an important part of a robust
nuclear structure program. Work along these lines is proceeding and is partly
presented in Ref.~\cite{Anderson:2010aq}. Extending beyond $A=2$ will be covered
in a forthcoming paper.

\begin{figure}[thb]
\includegraphics*[height=5cm]{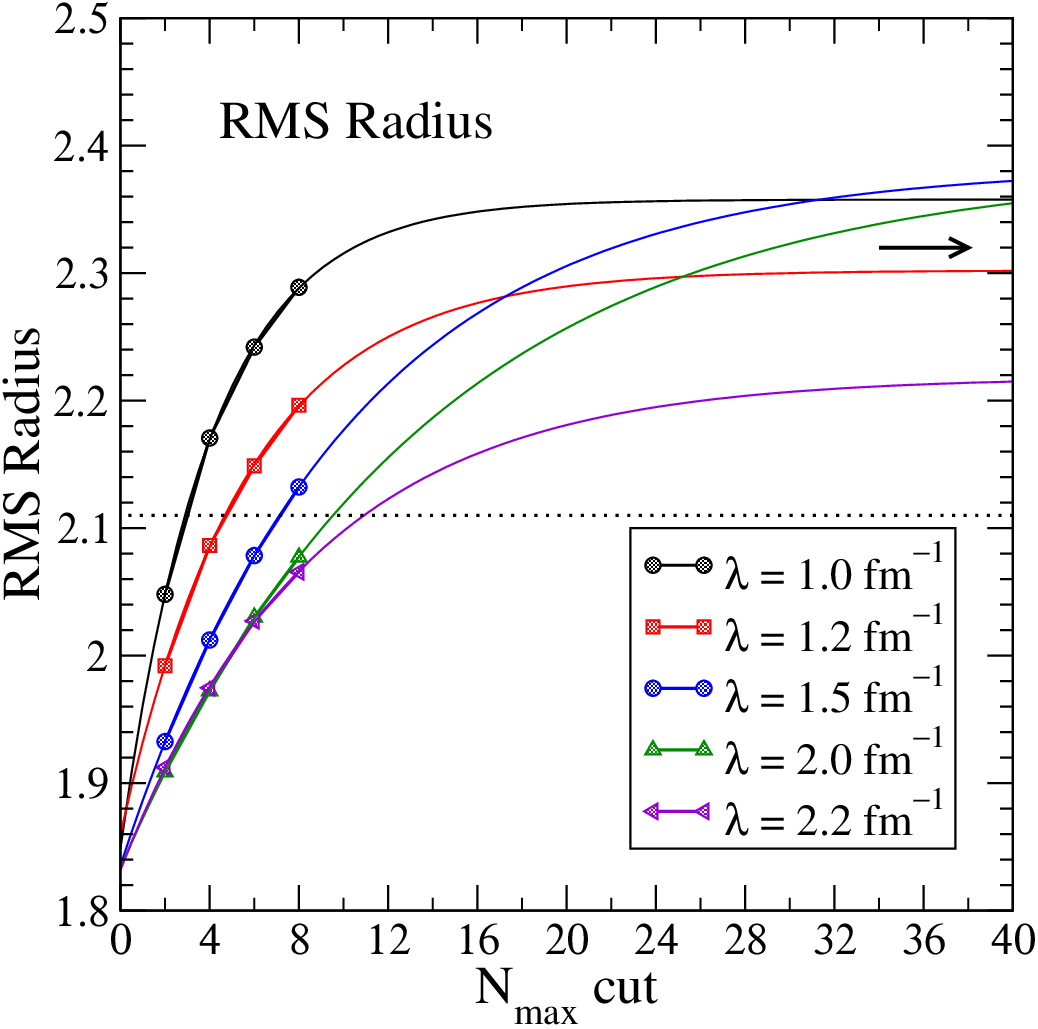}
\includegraphics*[height=5cm]{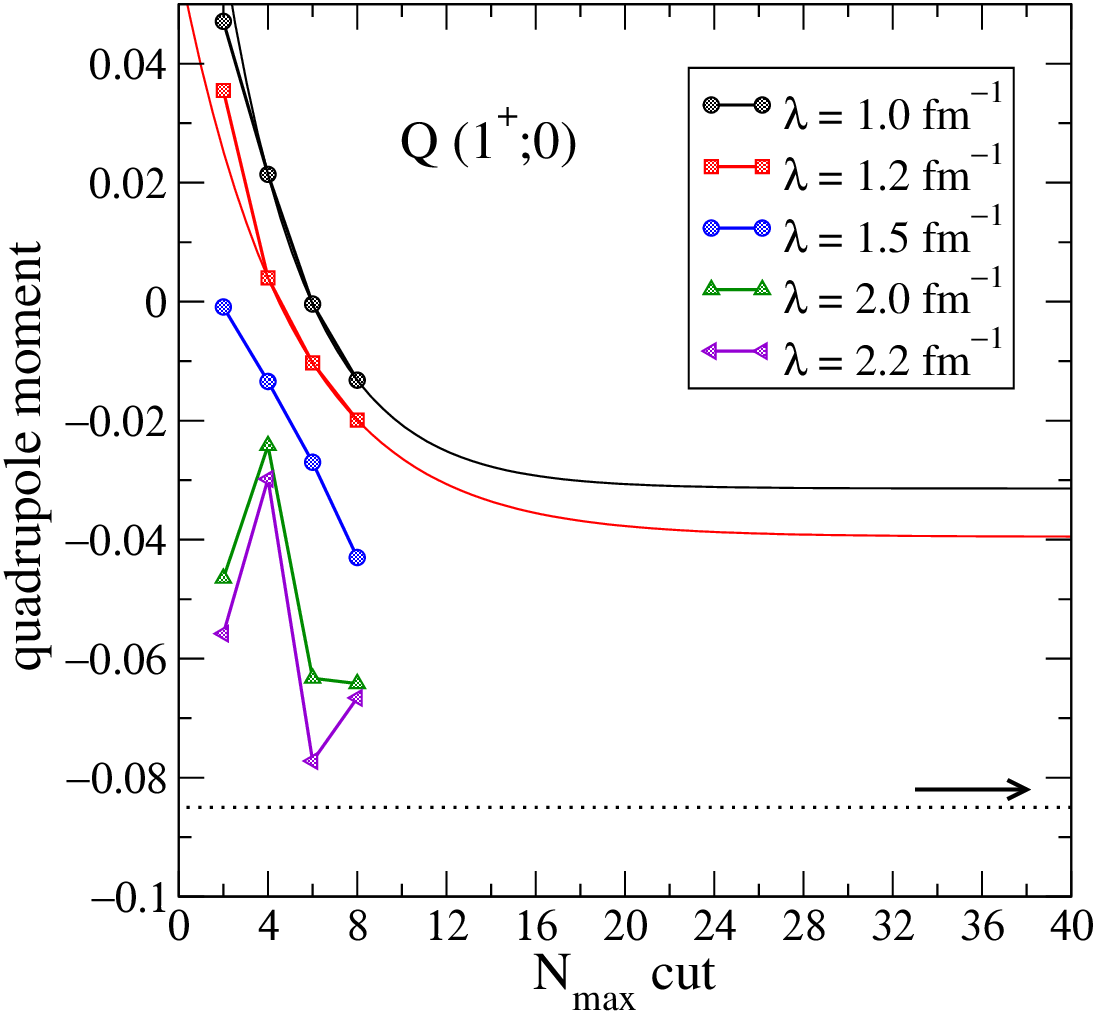}
\includegraphics*[height=5cm]{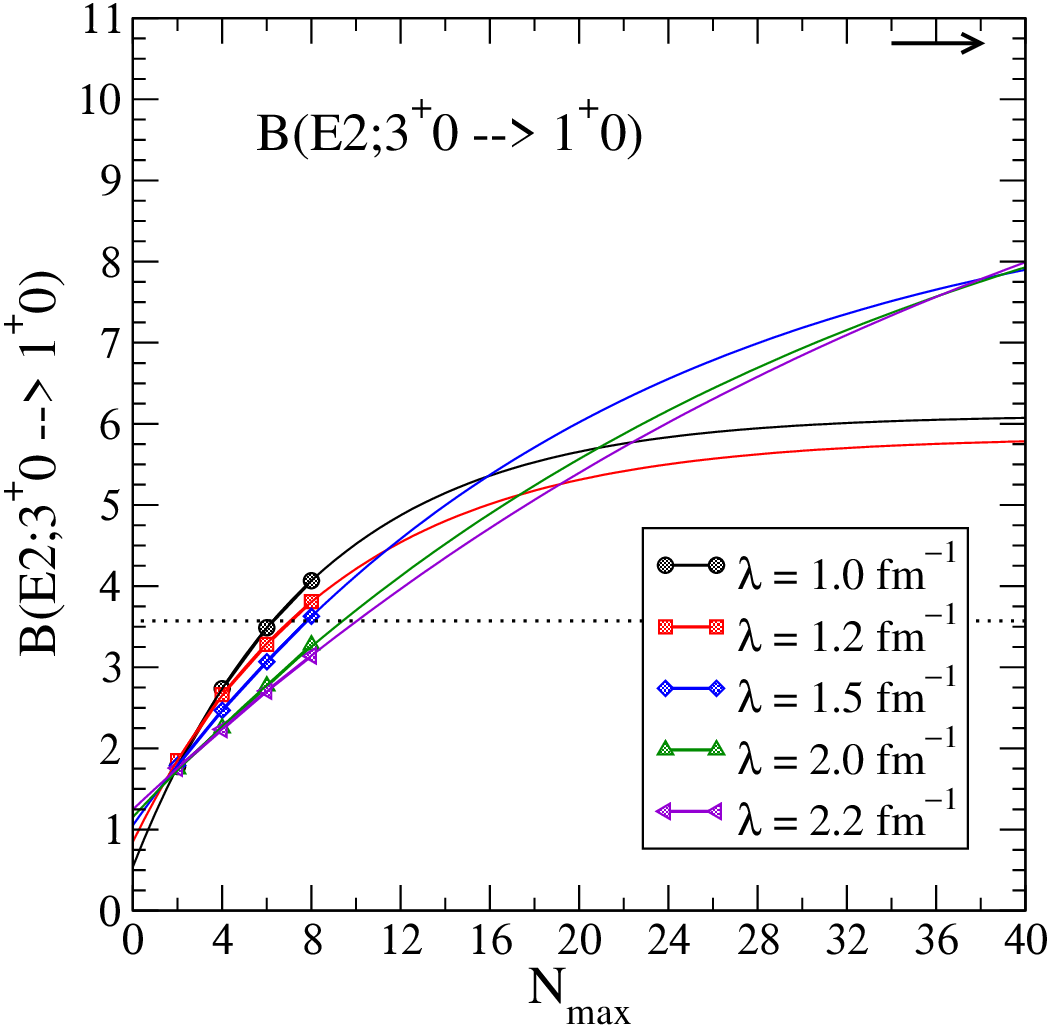}
\includegraphics*[height=5cm]{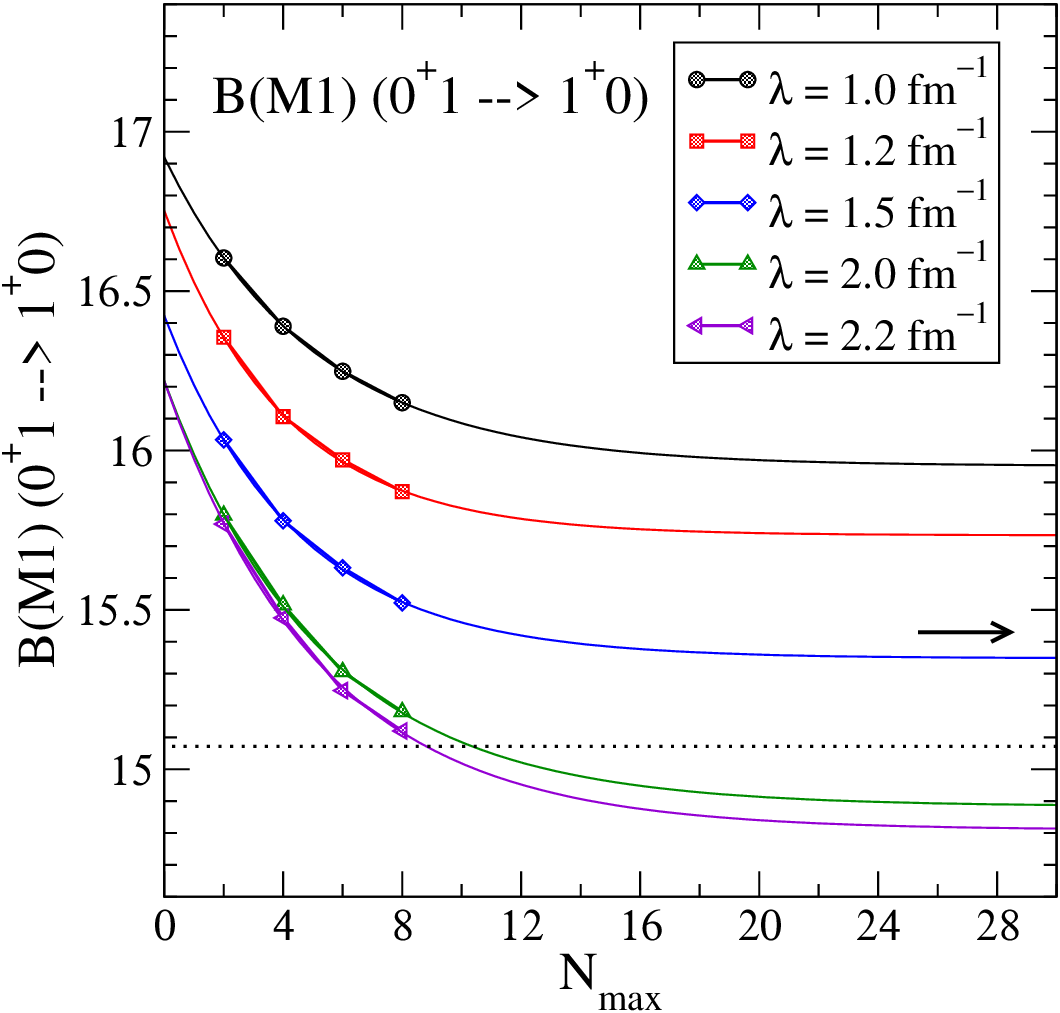}
\includegraphics*[height=5cm]{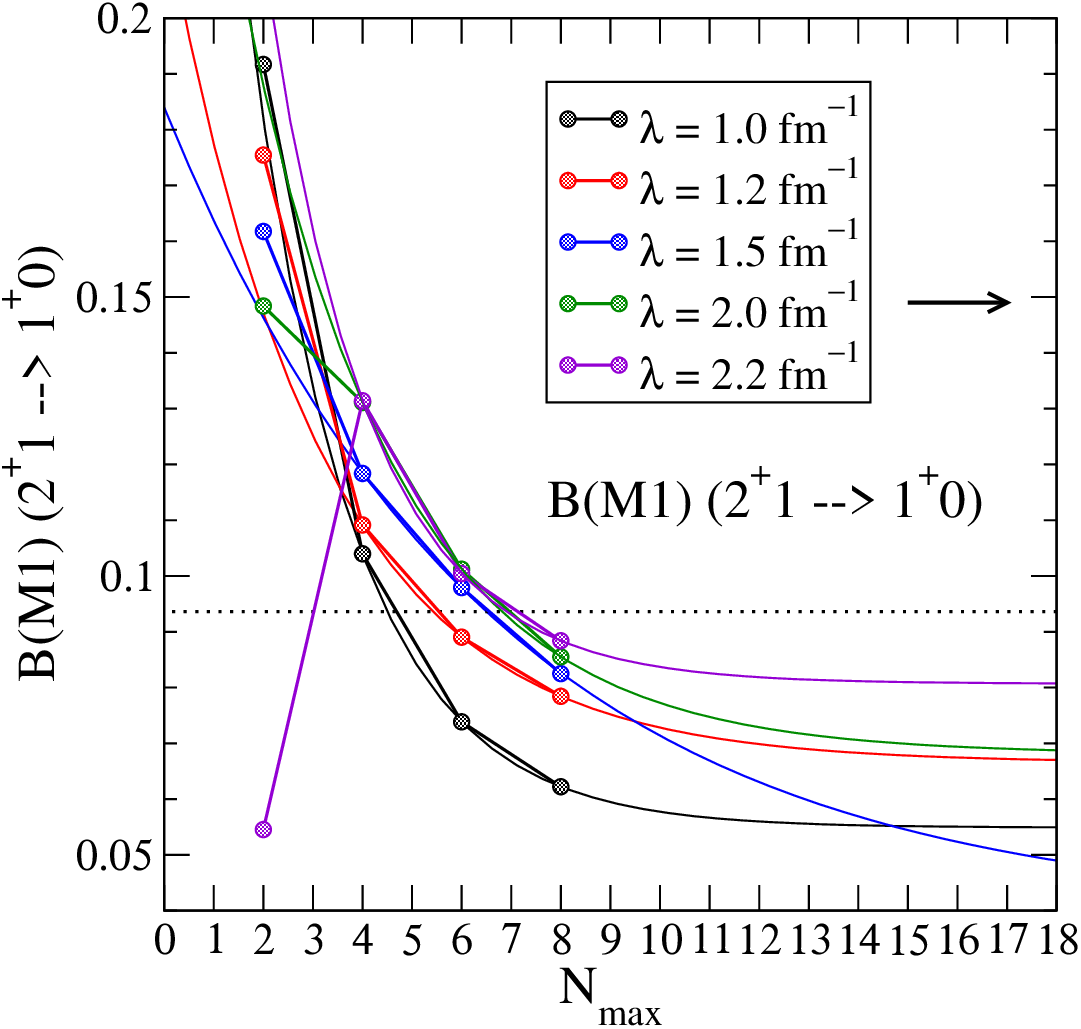}
\caption{(Color online) Various observables in $^6$Li as a function of \nmax\
for select $\lambda$'s. These results are with included initial NNN forces and
\atw\ = 300 and \ath\ = 40 and \hw\ = 20 MeV. The arrow shows the experimental
value and the dotted line shows the best LS result.}
\label{fig:He4_observable_convergence}
\end{figure}

Shown in Fig.~\ref{fig:He4_observable_convergence} are selected observables for
$^6$Li as a function of \nmax\ up to \nmax\ = 8. Included are simple
extrapolation curves shown by the dotted lines extending from the data points.
Table~\ref{tab:li6_observ} shows the values for \nmax\ = 8 at select values of
$\lambda$. In all cases the extrapolated values compare well to the established
Lee-Suzuki based results, but show room for improvement with respect to the
experimental values. Note the small scales on some of the plots, especially the
quadrupole moment and B(M1;$2^+1 \rightarrow 1^+0$).

Some of these observables exhibit a non-variational pattern in \nmax, such as the
quadrupole moment and the B(M1) shown here. These operators have strong coupling
between shells of \nmax and \nmax+2 that result in complex cancellations from
one truncation to another. However, SRG evolution seems to improve the
variational properties of these observables. Access to larger \nmax\ model
spaces will facilitate further study of these quantities.

\begin{table*}[tbh-]
\caption{\label{tab:li6_observ} Results for selected properties of
$^6$Li. Here we choose $\lambda = 1.0$ and $1.2 \fmi$ due to their convergence
properties. These results were obtained in a
basis space with \nmax\ = 8 and \hw\ = 20. All
results have included initial 3N forces at N$^2$LO. The LS results were obtained
at \nmax\ = 8 with \hw\ = 13 MeV~\cite{Navratil:2007we}. Note that \hw\ values for LS
and SRG procedures do not necessarily correspond to one another.}
\begin{center}
\begin{tabular}{|c||c|c|c|c|}
\hline
 Observable	  & $\lambda = 1.0$  & $\lambda = 1.2$ & Expt.		& LS  	  \\  	 
\hline
\hline
$r_p$ [fm]                            & 2.2841  &  2.1913 &	  2.32(3)  & 2.110	 \\		  
$Q (1_1^+ 0)$ [$e$ fm$^2$]            & $-$0.0132 & $-$0.0199 &  $-$0.082(2) & $-$0.085  \\		  
B(E2;$3_1^+ 0 \rightarrow 1_1^+ 0$)   & 4.0663  &  3.8087 &  10.69(84) & 3.5725  \\		  
B(M1;$0_1^+ 1 \rightarrow 1_1^+ 0$)   & 16.1499 & 15.8706 &  15.43(32) & 15.0717 \\		  
B(M1;$2_1^+ 1 \rightarrow 1_1^+ 0$)   & 0.0622  &  0.0784 &  0.149(27) & 0.0936  \\		  
\hline
\end{tabular}
\end{center}
\end{table*}

\section{Conclusions}
\label{sec:conclusions}


We have presented \abinit calculations of several light nuclei using SRG-evolved
three-nucleon forces. The results have  smooth convergence qualities with
respect to basis size, which enable reliable extrapolations. The extrapolated
(and converged where available) values are within the error bars of the best
existing Lee-Suzuki based calculations. Investigating the $\lambda$ dependence
of induced many-body forces, we find that they do not grow substantially as
$\lambda$ is lowered and the range of these effects is within the established LS
error bars. Analyzing the mechanism of flow for many-body terms reveals that the
SRG is driven by the natural hierarchy of the initial Hamiltonian and that it
preserves this hierarchy during evolution. This is qualitatively consistent with
studies of the same in one-dimension. Finally we present some first results of
various observables using SRG evolved many-body wavefunctions.


Our results here have focused mainly on $^6$Li observables and analysis in the
$A$ = 3 and 4 sectors. However, the input Hamiltonian files produced for this
work are universally valid for further calculations in the p-shell nuclei. Here,
we were limited in basis size (to \nmax\ = 8 in $^6$Li), but plan to apply the
evolved potentials at larger $A$ using codes capable of larger basis sizes. We
are first interested in studies of $^8$Be, $^{10}$B, and $^{12}$C, but this list
will undoubtedly expand. Also, we hope these potentials will be applied using
coupled cluster methods for even larger $A$~\cite{Hagen:2007hi}, and look
forward to applications of SRG evolution to external operators. Our work here
provides no indications of problems as high as $^6$Li with $T_{rel}$ as the SRG
generator. Other forms of the SRG generator may be useful in controlling the
growth of many-body forces in other nuclei.


In addition to the above ongoing work, we will apply the evolved
three-body interactions developed here to NCSM/RGM calculations~\cite{NCSM_RGM}
of light nuclear reactions. The NN-only evolved interactions have so far produced
good scattering and reaction results for s- and light p-shell nuclei. Adding the
evolved three-body interaction to the NCSM/RGM formalism will further improve
accuracy and allow us to extend its applicability to heavier p-shell and light
sd-shell nuclei. 

\begin{acknowledgments}

We thank E.~Anderson, E.~Ormand, R.~Perry, and S.~Quaglioni for useful comments.
This work was supported in part by the National Science Foundation under Grant
No.~PHY--0653312 and the UNEDF SciDAC Collaboration under DOE Grant
DE-FC02-07ER41457. This work was performed under the auspices of the U.S.
Department of Energy by Lawrence Livermore Laboratory under Contract
DE-AC52-07NA27344.

\end{acknowledgments}


\end{document}